\documentclass[titlepage,11pt]{article}
\usepackage{latexsym}
\usepackage{amsfonts}
\usepackage{amsmath}
\newcommand\blackslug{\hbox{\hskip 1pt \vrule width 4pt height 8pt depth 1.5pt
        \hskip 1pt}}
\newcommand\bbox{\hfill \quad \blackslug \bigbreak}
\def\d{\hbox{-}}

\def\l{,\ldots,}

\title{Edge-colouring eight-regular planar graphs}
\author{Maria Chudnovsky\thanks{Supported by NSF grants DMS-1001091 and IIS-1117631.}\\
Columbia University, New York, NY 10027
\\
\\
Katherine Edwards\thanks{Supported by an NSERC PGS-D3 Fellowship and a Gordon Wu Fellowship.}, Paul Seymour\thanks{Supported by ONR grant N00014-10-1-0680 and NSF grant DMS-0901075.}\\
Princeton University, Princeton, NJ 08544}

\date{January 13, 2012; revised \today}

\newtheorem{thm}{}[section]

\newcommand{\Proof}{\noindent{\bf Proof.}\ \ }
\newcommand{\indentitem}{\setlength\itemindent{25pt}}

\begin{document}
\maketitle
\begin{abstract}
It was conjectured by the third author in about 1973 that every $d$-regular 
planar graph (possibly with parallel edges) 
can be $d$-edge-coloured, provided that for every odd set $X$
of vertices, there are at least $d$ edges between $X$ and its complement.
For $d = 3$ this is the four-colour theorem, and the conjecture has been 
proved for all $d\le 7$, by various authors. Here we prove it for $d = 8$.
\end{abstract}

\section{Introduction}
One form of the four-colour theorem, due to Tait~\cite{tait}, asserts that a
$3$-regular planar graph can be $3$-edge-coloured if and only if it has no 
cut-edge. But when can $d$-regular planar graphs be $d$-edge-coloured?

Let $G$ be a graph. (Graphs in this paper are finite, and may have loops or parallel edges.)
If $X\subseteq V(G)$, $\delta_G(X) = \delta(X)$ denotes the set of all
edges of $G$ with an end in $X$ and an end in $V(G)\setminus X$.
We say that $G$ is {\em oddly $d$-edge-connected} if $|\delta(X)|\ge d$
for all odd subsets $X$ of $V(G)$. Since every perfect matching contains 
an edge of $\delta(X)$ for every odd set $X\subseteq V(G)$, it follows
that every $d$-regular $d$-edge-colourable graph is oddly $d$-edge-connected.
(Note that for a $3$-regular graph, being oddly $3$-edge-connected is 
the same as having no cut-edge, 
because if $X\subseteq V(G)$, then $|\delta(X)| = 1$ if and only if 
$|X|$ is odd and $|\delta(X)|<3$.) The converse is false, even for $d = 3$
(the Petersen graph is a counterexample); but for planar graphs perhaps the
converse is true. That is the content of the following 
conjecture~\cite{seymour}, proposed by the third author in about 1973.

\begin{thm}\label{mainconj}
{\bf Conjecture.} If $G$ is a $d$-regular planar graph, then $G$ is 
$d$-edge-colourable if and only if $G$ is oddly $d$-edge-connected.
\end{thm}

Some special cases of this conjecture have been proved. 
\begin{itemize}
\item For $d = 3$ 
it is the four-colour theorem, and was
proved by Appel and Haken~\cite{appelhaken1,appelhaken2,rsst}; 
\item for $d = 4,5$ it was proved by Guenin~\cite{guenin}; 
\item for $d = 6$ it was proved by Dvorak, Kawarabayashi
and Kral~\cite{dvorak}; 
\item for $d = 7$ it was proved by Kawarabayashi and the second 
author, and appears in the Master's thesis~\cite{katie} of the latter. The methods of the present paper can also be applied to the $d=7$ case, resulting in a proof somewhat simpler
than the original, and this simplified proof for the $d=7$ case will be presented in another, four-author paper~\cite{kawa}. 
\end{itemize}
Here we prove the next case, namely:

\begin{thm}\label{mainthm}
Every $8$-regular oddly $8$-edge-connected planar graph is $8$-edge-colourable.
\end{thm}
All these proofs (for $d>3$), including ours, proceed by induction on $d$. 
Thus we need to assume the truth of the result for $d = 7$. 

\section{An unavoidable list of reducible configurations.}

The graph we wish to edge-colour has parallel edges, but it is more convenient to work with the underlying simple graph.
If $H$ is $d$-regular and oddly $d$-edge-connected, then $H$ has no loops, because
for every vertex $v$, $v$ has degree $d$, and yet $|\delta_H(v)|\ge d$. (We write $\delta(v)$ for $\delta(\{v\})$.)
Thus to recover $H$ from the underlying simple graph $G$ say, we just need to know the number $m(e)$ of parallel edges
of $H$ that correspond to each edge $e$ of $G$. Let us say a {\em $d$-target} is a pair $(G,m)$
with the following properties
(where for $F\subseteq E(G)$, $m(F)$ denotes $\sum_{e\in F} m(e)$):
\begin{itemize}
\item $G$ is a simple graph drawn in the plane;
\item $m(e)\ge 0$ is an integer for each edge $e$;
\item $m(\delta(v)) = d$ for every vertex $v$; and
\item $m(\delta(X))\ge d$ for every odd subset $X\subseteq V(G)$.
\end{itemize}

In this language, \ref{mainconj} says that for every $d$-target $(G,m)$, there is a list of $d$ perfect matchings of $G$
such that every edge $e$ of $G$ is in exactly $m(e)$ of them. 
(The elements of a list need not be distinct.) If there is such a list we call it a {\em $d$-edge-colouring}, and say that $(G,m)$ is {\em
$d$-edge-colourable}.
For an edge $e\in E(G)$, we call $m(e)$ the {\em multiplicity} of $e$.
If $X\subseteq V(G)$, $G|X$ denotes the subgraph of $G$ induced on $X$.
We need:

\begin{thm}\label{counterex}
Let $(G,m)$ be a $d$-target, that is not $d$-edge-colourable, but such that every $d$-target with fewer vertices is $d$-edge-colourable. Then 
\begin{itemize}
\item $|V(G)|\ge 6$;
\item for every $X\subseteq V(G)$ with $|X|$ odd, if $|X|,|V(G)\setminus X|\ne 1$ then $m(\delta(X))\ge d+2$; and
\item $G$ is three-connected, and $m(e)\le d-2$ for every edge $e$.
\end{itemize}
\end{thm}
\Proof If $m(e) = 0$ for some edge $e$, we may delete $e$ without affecting the problem; so we may assume that $m(e)>0$ for every edge $e$.
It is easy to check that $G$ is connected and $|V(G)|\ge 6$ and we omit it.
For the second assertion
let $X\subseteq V(G)$ with $|X|$ odd and with $|X|,|V(G)\setminus X|\ne 1$. Thus $m(\delta(X))\ge d$ since $(G,m)$ is a $d$-target; suppose that $m(\delta(X))= d$.
There is a component of $G|X$ with an odd number of vertices, with vertex set $X'$ say; and so $m(\delta(X'))\ge d$ since
$(G,m)$ is a $d$-target. But $\delta(X')\subseteq \delta(X)$, and $m(e)>0$ for every edge $e$; and so $\delta(X') = \delta(X)$.
Since $G$ is connected it follows that $X' = X$, and so $G|X$ is connected. Similarly $G|Y$ is connected, where $Y = V(G)\setminus X$.
Replace each edge $e$ of $G$ by $m(e)$ parallel edges, forming $H$; and contract all edges of $H|Y$, forming a $d$-regular oddly $d$-edge-connected
planar graph $H_1$ with fewer vertices than $H$ (because $|Y|>1$). By hypothesis it follows that $H_1$ is $d$-edge-colourable. Similarly so is
the graph obtained from $H$ by contracting all edges of $G|X$. But these colourings can be combined to give a $d$-edge-colouring of $H$,
a contradiction. This proves that $m(\delta(X))> d$.  Since $m(\delta(v)) = d$ for every vertex $v$,
it follows that $m(\delta(X))$ has the same parity as $d|X|$, and so $m(\delta(X))\ge d+2$.
This proves the second assertion.

For the third assertion, suppose that $G$ is not three-connected.
Since $|V(G)|>3$, there is a partition $(X,Y,Z)$ of $V(G)$ where $X,Y\ne \emptyset$, with $|Z| = 2$, such that there are
no edges between $X$ and $Y$. Let $Z = \{z_1,z_2\}$ say. Either both $|X|,|Y|$ are odd, or they are both even. If they are both odd, then
since $\delta(X), \delta(Y)$ are disjoint subsets of $\delta(z_1)\cup \delta(z_2)$, and
$$m(\delta(X)), m(\delta(Y))\ge d = m(\delta(z_1)),m(\delta(z_2)),$$
we have equality throughout, and in particular $m(\delta(X)), m(\delta(Y)) = d$. But then $|X| = |Y| = 1$ from the second assertion,
contradicting that $|V(G)|\ge 6$. Now assume $|X|,|Y|$ are both even. Since $\delta(X\cup \{z_1\}), \delta(Y\cup \{z_2\})$ have the same union
and intersection as $\delta(z_1),\delta(z_2)$, it follows that $m(\delta(X\cup \{z_1\})) = d$, contrary to the second assertion.
Thus $G$ is three-connected.
Since $m(e)\ge 1$ for every edge $e$,
and $m(\delta(v)) = d$ for every vertex $v$, it follows that $m(e)\le d-2$ for every edge $e$. This proves the third assertion, and hence proves \ref{counterex}.~\bbox

A {\em triangle} is a region of $G$ incident with exactly three edges. If a triangle is incident with vertices $u,v,w$, for convenience we refer to it as
$uvw$, and in the same way an edge with ends $u,v$ is called $uv$.
Two edges are {\em disjoint} if they are distinct and no vertex is an end of both of them, and otherwise they {\em meet}.
Let $r$ be a region of $G$, and let $e\in E(G)$ be incident with $r$; let $r'$ be the other region incident with $e$. We say that $e$
is {\em $i$-heavy} (for $r$), where $i\ge 2$, if either
$m(e)\ge i$ or $r'$ is a triangle $uvw$ where $e = uv$ and
$$m(uv)+\min(m(uw),m(vw))\ge i.$$
We say $e$ is a {\em door} for $r$ if $m(e) = 1$ and there is an edge $f$ incident with $r'$ and disjoint from $e$ with $m(f) = 1$.
We say that $r$ is {\em big} if there are at least four doors for $r$, and {\em small} otherwise. A {\em square} is a region with length four.

Since $G$ is drawn in the plane and is two-connected, every region $r$ has boundary some cycle which we denote by $C_r$.
In what follows we will be studying cases in which certain configurations of regions are present in $G$.
We will give a list of regions the closure of the union of which is a disc. For convenience, for an edge $e$
in the boundary of this disc, we call the region outside the disc incident with $e$ the ``second region'' for $e$; and we write $m^+(e) = m(e)$ if the second region
is big, and $m^+(e) = m(e)+1$ if the second region is small. This notation thus depends not just on $(G,m)$ but on what regions we have specified, so it is
imprecise, and when there is a danger of ambiguity we will specify it more clearly.

Let us say an $8$-target $(G,m)$ is {\em prime} if 
\begin{itemize}
\item $m(e)>0$ for every edge $e$;
\item $|V(G)|\ge 6$;
\item $m(\delta(X))\ge 10$ for every $X\subseteq V(G)$ with $|X|$ odd and $|X|,|V(G)\setminus X|\ne 1$;
\item $G$ is three-connected, and $m(e)\le 6$ for every edge $e$;
\end{itemize}
and in addition $(G,m)$ contains none of of the following:
\begin{itemize}
{\indentitem\item[{\bf Conf(1):}] A triangle $uvw$ where $u,v$ both have degree three.}
{\indentitem\item[{\bf Conf(2):}] A triangle $uvw$, where $u$ has degree three and its third neighbour $x$ satisfies 
$$m(ux)<m(uw)+m(vw).$$}
{\indentitem\item[{\bf Conf(3):}] Two triangles $uvw,uwx$ with $m(uv) + m(uw) + m(vw)+ m(ux)\ge 8$.}
{\indentitem\item[{\bf Conf(4):}] A square $uvwx$ where $m(uv)+m(vw)+m(ux)\ge 8$ and 
$$(m(uv),m(vw),m(wx),m(ux))\ne (4,2,1,2).$$}
{\indentitem\item[{\bf Conf(5):}] Two triangles $uvw,uwx$ where $m^+(uv)+m(uw)+m^+(wx) \ge 7$.}
{\indentitem\item[{\bf Conf(6):}] A square $uvwx$ where $m^+(uv)+m^+(wx)\ge 7$.}
{\indentitem\item[{\bf Conf(7):}] A triangle $uvw$ with $m^+(uv)+m^+(uw) \ge 7$.}
{\indentitem\item[{\bf Conf(8):}] A triangle $uvw$, where $m(uv) = 3$, $m(uw) = 2$, $m(vw) = 2$, and the second region for one of $uv,uw,vw$ has no door disjoint from $uvw$.}
{\indentitem\item[{\bf Conf(9):}] A triangle $uvw$ with $m(uv), m(uw), m(vw) = 2$, such that $u$ has degree at least four, and the second regions for $uv,uw$ 
both have at most one door, and no door that is disjoint from $uvw$.}
{\indentitem\item[{\bf Conf(10):}] A square $uvwx$ and a triangle $wxy$, where $m(uv) = m(wx)=m(xy)=2$, and $m(vw) = 4$.}
{\indentitem\item[{\bf Conf(11):}] A square $uvwx$ and a triangle $wxy$, where $m(uv) \ge 3$, $m(wy)\ge 3$, $m(wx) = 1$, $m(ux)\le 3$, and $m^+(xy)\ge 3$.}
{\indentitem\item[{\bf Conf(12):}] A square $uvwx$ and a triangle $wxy$, where $m^+(uv),m(vw)\ge 2$, $m(wx) = m(wy) = 2$, $m(ux)\le 3$, and $m^+(xy)\ge 3$.}
{\indentitem\item[{\bf Conf(13):}] A region $r$ of length five, with edges $e_1\l e_5$ in order, where $m(e_1)\ge \max(m(e_2), m(e_5))$, 
$m(e_1)+m(e_2)+m(e_3)\ge 8$ and $m^+(e_1)+m^+(e_4)\ge 7$.}
{\indentitem\item[{\bf Conf(14):}] A region $r$ and an edge $e$ of $C_r$, such that $m^+(e) \ge 6$ and at most six edges of $C_r$ disjoint from $e$ are doors for $r$.}
{\indentitem\item[{\bf Conf(15):}] A region $r$ with length at least four,
and an edge $e$ of $C_r$, such that $m^+(e)\ge 4$ and every edge of $C_r$ disjoint from $e$ is $3$-heavy.}
{\indentitem\item[{\bf Conf(16):}] A region $r$ and an edge $uv$ of $C_r$, and a triangle $uvw$,
such that $m(uv)+ m^+(uw)\ge 4$, and $m(vw)\le m(uw)$, and the second edge of $C_r$ incident with $u$ has multiplicity at most $m(uw)$,
and every edge of $C_r$ not incident with $u$ is $3$-heavy.}
{\indentitem\item[{\bf Conf(17):}] A region $r$ with length at least five, and an edge $e$ of $C_r$, such that $m^+(e)\ge 5$,
every edge $f$ of $C_r$ disjoint from $e$ satisfies $m^+(f)\ge 2$, and at most one of them is not $3$-heavy.}
{\indentitem\item[{\bf Conf(18):}] A region $r$ with length at least four and an edge $uv$ of $C_r$, and a triangle $uvw$,
such that $m^+(uw)+m(uv)\ge 5$, and $m(vw)\le m(uw)$, and the second edge of $C_r$ incident with $u$ has multiplicity at most $m(uw)$,
and either
\begin{itemize}
\item $m(uv) = 3$ and $uv$ is $5$-heavy, and every edge $f$ of $C_r$ disjoint from $uv$ satisfies $m^+(f)\ge 2$, and at most one of them is not $3$-heavy, or
\item $m^+(f)\ge 2$ for every edge $f$ of $C_r$ not incident with $u$, and at most one such edge is not $3$-heavy.
\end{itemize}}
{\indentitem\item[{\bf Conf(19):}] A region $r$ with length at least five and an edge $e$ of $C_r$, such that $m^+(e)\ge 5$,
every edge of $C_r$ disjoint from $e$ is $2$-heavy, and at most two of them are not $3$-heavy.}
\end{itemize}

We will prove these restrictions are too much, that in fact no $8$-target is prime (theorem~\ref{unav}). 
To deduce \ref{mainthm}, we will show that if there is a counterexample, then some counterexample is prime; but
for this purpose, just choosing a counterexample with the minimum number of vertices is not enough, and we need a more delicate minimization.
If $(G,m)$ is a $d$-target, its {\em score sequence} is the $(d+1)$-tuple $(n_0,n_1 \l n_d)$ where $n_i$ is the number of edges
$e$ of $G$ with $m(e) = i$. If $(G,m)$ and $(G',m')$ are $d$-targets, with score sequences $(n_0 \l n_d)$ and $(n_0' \l n_d')$ respectively,
we say that $(G',m')$ is {\em smaller} than $(G,m)$ if either
\begin{itemize}
\item $|V(G')|< |V(G)|$, or
\item $|V(G')|= |V(G)|$ and there exists $i$ with $1\le i\le d$ such that $n_i'>n_i$, and $n_j' = n_j$ for all $j$ with $i<j\le d$, or
\item $|V(G')|= |V(G)|$, and $n_j' = n_j$ for all $j$ with $0<j\le d$, and $n_0'<n_0$.
\end{itemize}
(The anomalous treatment of $n_0$ is just a device to allow $d$-targets to have edges with $m(e) = 0$, while minimum $d$-counterexamples have none.)
If some $d$-target is not $d$-edge-colourable, then we can choose a $d$-target $(G,m)$ with the following properties:
\begin{itemize}
\item $(G,m)$ is not $d$-edge-colourable
\item every smaller $d$-target is $d$-edge-colourable.
\end{itemize}
Let us call such a pair $(G,m)$ a {\em minimum $d$-counterexample}. 
To prove \ref{mainthm}, we prove two things:
\begin{itemize}
\item No $8$-target is prime (theorem~\ref{unav}), and
\item Every minimum $8$-counterexample is prime (theorem~\ref{reduc}).
\end{itemize}
It will follow that there is no minimum $8$-counterexample, and so the theorem is true.

\section{Discharging and unavoidability}

In this section we prove the following, with a discharging argument.

\begin{thm}\label{unav}
No $8$-target is prime.
\end{thm}

The proof is broken into several steps, through this section.
Let $(G,m)$ be a $8$-target, where $G$ is three-connected. For every region $r$, we define 
$$\alpha(r) = 8 - 4|E(C_r)| +  \sum_{e\in E(C_r)}m(e).$$
We observe first:

\begin{thm}\label{totalsum}
The sum of $\alpha(r)$ over all regions $r$ is positive.
\end{thm}
\Proof
Since $(G,m)$ is a $8$-target, $m(\delta(v)) = 8$ for each vertex $v$, and, summing over all $v$, we deduce that
$2m(E(G)) = 8|V(G)|$. By Euler's formula, the number $R$ of regions of $G$ satisfies $|V(G)| - |E(G)| + R = 2$,
and so $2m(E(G)) - 8|E(G)| + 8R = 16$.
But $2m(E(G))$ is the sum over all regions $r$, of  $\sum_{e\in E(C_r)}m(e)$, and $8R-8|E(G)|$ is the sum over all regions $r$ of $8-4|E(C_r)|$.
It follows that the sum of $\alpha(r)$ over all regions $r$ equals $16$. This proves \ref{totalsum}.~\bbox

We normally wish to pass one unit of charge from every small region to every big region with which it shares an edge; except that in some
rare circumstances, sending one unit is too much, and we only send $1/2$ or $0$.
More precisely, for every edge $e$ of $G$, define $\beta_e(s)$ for each region $s$ as follows.
Let $r,r'$ be the two regions incident with $e$. 
\begin{itemize}
\item If $s\ne r,r'$ then $\beta_e(s) = 0$.
\item If $r,r'$ are both big or both small then  $\beta_e(r), \beta_e(r') = 0$. 
\end{itemize}
Henceforth we assume that $r$ is big and $r'$ is small; let 
$f,f'$ be the edges of $C_r\setminus e$ that share an end with $e$.
\begin{itemize}
	\item[{\bf 1:}] If $e$ is a door for $r$ (and hence $m(e) = 1$) then $\beta_e(r) = \beta_e(r') = 0$.
	\item[{\bf 2:}] If $m(e)= 2$ and $m^+(f) = m^+(f') = 6$ then $\beta_e(r) = \beta_e(r') =0$. 
	\item[{\bf 3:}] If $m(e) = 2$ and $m^+(f) = 6$ and $m^+(f') = 5$ or vice versa then $\beta_e(r) = -\beta_e(r') =1/2$.
	\item[{\bf 4:}] If $m(e) = 3$ and $m^+(f) = m^+(f') = 5$ then $\beta_e(r) = \beta_e(r') =0$.
	\item[{\bf 5:}] If $m(e) = 3$ and exactly one of $m^+(f), m^+(f') = 5$, then $\beta_e(r) = -\beta_e(r') =1/2$.
	\item[{\bf 6:}] Otherwise $\beta_e(r) = -\beta_e(r') = 1$.
\end{itemize}
(Think of $\beta_e$ as passing some amount of charge between the two regions incident with $e$.)
For each region $r$, define $\beta(r)$ to be the sum of $\beta_e(r)$ over all edges $e$. 
We see that the sum of $\beta(r)$ over all regions $r$ is zero. 

The effect of $\beta$ is passing charge from small regions to big regions with which they share an edge.
We need another ``discharging'' function, that passes charge from triangles to small regions with which they share an edge.
If $r$ is a triangle, incident with edges $e,f,g$, we define its {\em multiplicity} $m(r) = m(e) + m(f) + m(g)$.
A region $r$ is {\em tough} if $r$ is a triangle, its multiplicity is at least five, and if $r = uvw$ where $m(uv) = 1$ and $m(uw) = m(vw) = 2$, then
$m^+(uw)+m^+(vw)\ge 5$.
For every edge $e$ of $G$, define $\gamma_e(s)$ for each region $s$ as follows.
Let $r,r'$ be the two regions incident with $e$.
\begin{itemize}
\item If $s\ne r,r'$ then $\gamma_e(s) = 0$.
\item If one of $r,r'$ is big, or neither is tough, or they both are tough,
then $\gamma_e(r) = \gamma_e(r') = 0$. 
\end{itemize}
Henceforth we assume that $r'$  is tough, and $r$ is small and not tough.
Let $e,e_1,e_2$ be the edges incident with $r'$, and let $r_1,r_2$ be the regions different from $r'$ incident with $e_1,e_2$ respectively.
\begin{itemize}
\item[{\bf 1:}] If $m(e) = 1$ and $m(e_1), m(e_2)\ge 2$, and $m^+(e_1)+m^+(e_2)\ge 6$ then $\gamma_e(r) = -\gamma_e(r') = 1$.
\item[{\bf 2:}] If $m(e) = 1$ and $m^+(e_1)\ge 4$ and $m(e_2) = 1$ and $r_2$ is small, then $\gamma_e(r) = -\gamma_e(r') = 1/2$.
\item[{\bf 3:}] If $m(e) = 1$ and $m(e_1) = 3$ and $m(e_2) = 1$ and $r_2$ is small, and the edge $f$ of $C_{r}\setminus e$ that shares an end
with $e,e_1$ satisfies $m(f) = 4$, then $\gamma_e(r) = -\gamma_e(r') = 1/2$.
\item[{\bf 4:}] If $m(e) = 2$ and $m(e_1), m(e_2)\ge 2$ and $m^+(e_1)+m^+(e_2)\ge 5$, and either 
\begin{itemize}
\item $r$ has more than one door, or 
\item some door for $r$ is disjoint from $e$, or
\item some edge $f$ of $C_r$ consecutive with $e$ has multiplicity four, and $r_1,r_2$ are both small,
\end{itemize}
then
$\gamma_e(r) = -\gamma_e(r') = 1$.
\item[{\bf 5:}] If $m(e) = 2$ and $m(e_1), m(e_2)= 2$ and some end of $e$ has degree three, 
incident with $e_1$ say, and $r_1$ is small and $r_2$ is big, then
$\gamma_e(r) = -\gamma_e(r') = 1/2$.
\item[{\bf 6:}] If $m(e) = 3$ and $m(e_1),m(e_2)= 2$ then $\gamma_e(r) = -\gamma_e(r') = 1$.
\item[{\bf 7:}] Otherwise $\gamma_e(r) = \gamma_e(r') = 0$.
\end{itemize}
For each region $r$, define $\gamma(r)$ to be the sum of $\gamma_e(r)$ over all edges $e$.
Again, the sum of $\gamma(r)$ over all regions $r$ is zero.

We observe that, immediately from the rules, we have

\begin{thm}\label{nonzero}
Let $e$ be incident with regions $r,r'$. Then $\beta_e(r)$ is non-zero only if exactly one of 
$r,r'$ is big; and $\gamma_e(r)$ is non-zero only if exactly one of $r,r'$ is tough and neither is big.
Thus in all cases, at most one of $\beta_e(r), \gamma_e(r)$ is non-zero. Moreover $|\beta_e(r)+\gamma_e(r)|\le 1$.
\end{thm}

Let $\alpha, \beta, \gamma$ be as above. Then the sum over all regions $r$ of $\alpha(r)+\beta(r)+\gamma(r)$ is positive,
and so there is a region $r$ with $\alpha(r)+\beta(r)+\gamma(r)>0$. Let us examine the possibilities for such a region.  
There now begins a long case analysis, and to save writing we just say ``by Conf(7)'' instead of ``since $(G,m)$ does not contain Conf(7)'', and so on.

\begin{thm}\label{bigovercharge}
If $r$ is a big region and $\alpha(r)+\beta(r)+\gamma(r)>0$, then $(G,m)$ is not prime.
\end{thm}
\Proof
Suppose that $(G,m)$ is prime.
Let $C = C_r$. Since $r$ is big it follows that $\gamma(r) = 0$, and so $\alpha(r)+\beta(r)>0$; that is,
$$\sum_{e\in E(C)}(4-m(e) - \beta_e(r)) < 8.$$
For $e\in E(C)$, define $\phi(e) = m(e) + \beta_e(r)$, and let us say $e$ is {\em major} if $\phi(e)>4$.
If $e$ is major, then since $\beta_e(r)\le 1$, it follows that $m(e)\ge 4$; and so $\beta_e(r)$ is an integer, from the $\beta$-rules, and therefore $\phi(e) \ge 5$. 
Moreover, no two major edges are consecutive,
since $G$ has minimum degree at least three.

Let $D$ be the set of doors for $C$.
Let 
\begin{itemize}
\item $\xi = 1$ if there are consecutive edges $e,f$ in $C$ such that $\phi(e) >5$ and $f$ is a door for $r$
\item $\xi = 2$ if there is no such pair $e,f$.
\end{itemize}
(1) {\em Let $e,f,g$ be the edges of a path of $C$, in order, where $e,g$ are major. Then}
$$(4-\phi(e))+2(4-\phi(f))+(4-\phi(g)) \ge 2\xi|\{f\}\cap D|.$$
Let $r_1,r_2,r_3$ be the regions different from $r$ incident with $e,f,g$ respectively.
Now $m(e)\le 6$ since $(G,m)$ is prime, and if $m(e) = 6$ then $r_1$ is big, by Conf(14), and so $\beta_e(r) = 0$; and so in any case, $\phi(e)\le 6$.
Similarly $\phi(g)\le 6$. Also, $\phi(e), \phi(g)\ge 5$ since $e,g$ are major. Thus $\phi(e)+\phi(g)\in \{10,11,12\}$.

Suppose that $\phi(e)+\phi(g) = 12$. We must show that $\phi(f) \le 2-\xi|\{f\}\cap D|$. Now $m(e)\ge 5$, and so $m(f)\le 2$, since $G$ is three-connected. 
If $m(f) = 2$ then $f\notin D$, and $\beta_f(r) = 0$ from the $\beta$-rules; and so $\phi(f) \le 2-\xi |\{f\}\cap D|$.
If $m(f) = 1$, then $\beta_f(r)\le 1$, so we may assume that $f\in D$; but then $\xi = 1$ and $\phi(f) = 1\le  2-\xi |\{f\}\cap D|$.

Next suppose that $\phi(e)+\phi(g) = 11$. We must show that $\phi(f) \le 5/2-\xi|\{f\}\cap D|$.
Again one of $\phi(e),\phi(g)\ge 6$, say $\phi(e) = 6$; and so
$m^+(e)\ge 6$. In particular $m(e)\ge 5$, and so $m(f)\le 2$. Since $\phi(g)\ge 5$ we have $m^+(g)\ge 5$, and so if $m(f) = 2$, then $\beta_f(r) \le 1/2$ 
from the $\beta$-rules; and since $f\notin D$ we have $\phi(f) \le 5/2-\xi|\{f\}\cap D|$. If $m(f) = 1$, then $\phi(f)\le 2$, and so we may assume that
$f\in D$; but then $\xi = 1$ and $\phi(f) = 1$, and again $\phi(f) \le 5/2-\xi|\{f\}\cap D|$.

Finally, suppose that $\phi(e)+\phi(g) =10$. We must show that $\phi(f) \le 3-\xi|\{f\}\cap D|$. Suppose that $m(f) \ge 3$. Since $m^+(e), m^+(g)\ge 5$ (because $e,g$ are major),
it follows that $m(f) = 3$, and $m(e) = m(g) = 4$ because $G$ is three-connected; but then $\beta_f(r) = 0$ from the $\beta$-rules, and since $f\notin D$ we have
$\phi(f) \le 3-\xi|\{f\}\cap D|$. Next suppose that $m(f) = 2$. Then $\phi(f)\le 3 = 3-\xi|\{f\}\cap D|$ as required. 
Lastly if $m(f) = 1$, then $\phi(f)\le 2$, so we may assume that
$f\in D$; but then $\xi \le 2$ and $\phi(f) = 1 \le 3-\xi|\{f\}\cap D|$.
This proves (1).
\\
\\
(2) {\em Let $e,f$ be consecutive edges of $C$, where $e$ is major. Then }
$$(4-\phi(e))+2(4-\phi(f)) \ge 2\xi|\{f\}\cap D|.$$
We have $\phi(e)\in \{5,6\}$. Suppose that $\phi(e) = 6$. We must show that $\phi(f)\le 3-\xi|\{f\}\cap D|$; but $m(f)\le 2$ since $m(e)\ge 5$, and so
$\phi(f)\le 3$. We may therefore assume that $f\in D$; but then $\xi = 1$ and $\phi(f)=1\le 3-\xi|\{f\}\cap D|$. 
Next, suppose that $\phi(e) = 5$; then we must show that $\phi(f)\le 7/2-\xi|\{f\}\cap D|$. Since $m(e)\ge 4$, it follows that $m(f)\le 3$. If 
$m(f) = 3$ then $m^+(e) = 5$ and so $\beta_f(r)\le 1/2$,
from the $\beta$-rules; but then $\phi(f)\le 7/2-\xi|\{f\}\cap D|$. If $m(f)\le 2$, then $\phi(f)\le 1$, so we may assume that $f\in D$;
but $\xi\le 2$, and so $\phi(f)=1\le  7/2-\xi|\{f\}\cap D|$. This proves (2).

\bigskip

For $i = 0,1,2$, let $E_i$ be the set of edges $f\in E(C)$ such that $f$ is not major, and $f$ meets exactly $i$ major edges in $C$.
Let $D$ be the set of doors for $C$.
By (1), for each $f\in E_2$ we have 
$$\frac{1}{2}(4-\phi(e))+(4-\phi(f))+\frac{1}{2}(4-\phi(g))  \ge \xi|\{f\}\cap D|$$ 
where $e,g$ are the major edges meeting $f$. 
By (2), for each $f\in E_1$ we have
$$\frac{1}{2}(4-\phi(e))+(4-\phi(f)) \ge \xi|\{f\}\cap D|$$
where $e$ is the major edge consecutive with $f$.
Finally, for each $f\in E_0$ we have
$$4-\phi(f)\ge  \xi|\{f\}\cap D|$$
since $\phi(f)\le 4$, and $\phi(f) = 1$ if $f\in D$. Summing these inequalities over all $f\in E_0\cup E_1\cup E_2$, we deduce that
$\sum_{e\in E(C)}(4-\phi(e))\ge \xi|D|$.
Consequently 
$$8> \sum_{e\in E(C)}(4-m(e) - \beta_e(r)) \ge \xi |D|.$$ 
But $|D|\ge 4$ since $r$ is big, and so $\xi = 1$ and $|D|\le 7$, a contradiction by Conf(14).
This proves \ref{bigovercharge}.~\bbox

\begin{thm}\label{triovercharge1}
If $r$ is a triangle that is not tough, and $\alpha(r)+\beta(r)+\gamma(r)>0$, then $(G,m)$ is not prime.
\end{thm}
\Proof
Suppose $(G,m)$ is prime, and let $r = uvw$.
Suppose first that $r$ has multiplicity five; and hence,
since it is not tough, we may assume that $m(uv) = 1$ and $m(uw) = m(vw) = 2$, and
the second regions for $uw,vw$ are both big. Thus from the $\beta$-rules, $\beta_{uw}(r),\beta_{vw}(r)=-1$,
and since $\beta_{uv}(r) + \gamma_{uv}(r)\le 1$, we deduce that $\beta(r)+\gamma(r)\le -1$. But
$$\alpha(r) = - 4 +  m(uv) + m(vw) + m(uw) =1,$$
contradicting that $\alpha(r)+\beta(r)+\gamma(r)>0$. Thus $r$ has multiplicity at most four.

Since $\alpha(r) = - 4 +  m(uv) + m(vw) + m(uw)\le 0$, and $\beta(r)\le 0$,
it follows that $\gamma(r)>0$. Hence $\gamma_e(r)>0$ for some edge $e$ incident with $r$, say $e = uv$. 
\\
\\
(1) {\em $m(e) = 1$ for every edge $e$ incident with $r$ such that $\gamma_e(r)>0$.}
\\
\\
For suppose that $m(e)>1$ and $\gamma_e(r)>0$. Since $r$ has multiplicity at most four it follows that $m(e)= 2$.
Since  $\gamma_e(r)>0$, there is a vertex $x\ne w$ such that $uvx$ is a triangle, and
$m(ux),m(vx)\ge 2$, and one of $m^+(ux),m^+(vx)$ is at least three, say $m^+(ux) \ge 3$; and $r$ has two doors.
By Conf(5), $m^+(vw)=1$, and so $\beta_{vw}(r) = -1$ and $\beta_{uw}\le 0$, and hence
$\beta(r)\le -1$; yet $\gamma(r)\le 1$, contradicting that $\alpha(r)+\beta(r)+\gamma(r)>0$.
This proves (1).
\\
\\
(2) {\em There is no edge $e$ incident with $r$ and with a big region such that $m(e) = 1$.}
\\
\\
Let $r$ be incident with edges $e,f,g$, and suppose that $m(e) = 1$ and $e$ is incident with a big region. Thus $\beta(r)\le -1$, and so $\gamma(r)>1$; and consequently
$\gamma_f(r),\gamma_g(r)>0$, and therefore $m(f) = m(g)= 1$ from (1). But then $\alpha(r)=-1$, and yet $\gamma(r)\le 2$, contradicting that
$\alpha(r)+\beta(r)+\gamma(r)> 0$. This proves (2).

\bigskip

Choose $e$ with $\gamma_e(r)>0$, say $e = uv$. Thus $m(uv) = 1$, and
there is a tough triangle $r'=uvx$ say. 
By Conf(3), $r'$ has multiplicity at most six.
\\
\\
(3) {\em We may assume that $m^+(ux)\le 3$ and $m^+(vx)\le 3$.}
\\
\\
For suppose that $m^+(ux)\ge 4$. By (2), $m^+(vw)\ge 2$, contrary to Conf(5). This proves (3).

\bigskip

Now $\gamma_{uv}(r)>0$, and from
(1), (3), it follows that $\gamma_{uv}(r$ is determined by the first $\gamma$-rule. 
In particular, $m^+(ux)= 3$, and $m^+(vx)= 3$.
By Conf(16), $uw$ and $vw$ are not $3$-heavy, and so by the same argument $\gamma_{uw}(r)=0$ and $\gamma_{vw}(r)=0$; and so $\gamma(r) = 1$. 
Consequently $\alpha(r)>-1$, and so we may assume that $m(uw) = 2$. Let $r_1$ be the second region for $uw$. Now
$m(ux)+m(uv)+m(uw)\le 6$, and so there is an edge $f$ incident with $r_1$ and $u$ different from $uw,ux$. 
Moreover, $m(f)\le 3$, since $m(ux)+m(uv)+m(uw)\ge 5$; and so if $r_1$
is big then $\beta_{uw}(r) = -1$, a contradiction. Thus $r_1$ is small, contrary to 
Conf(5).
This proves \ref{triovercharge1}.~\bbox

\begin{thm}\label{triovercharge2}
If $r$ is a tough triangle with $\alpha(r)+\beta(r)+\gamma(r)>0$, then $(G,m)$ is not prime.
\end{thm}
\Proof
Suppose $(G,m)$ is prime, and let $r = uvw$.
Now $\alpha(r) = m(uv)+m(vw)+m(uw)-4$,  so 
$$ m(uv)+m(vw)+m(uw)+\beta(r) + \gamma(r)>4.$$
Let $r_1,r_2,r_3$ be the regions different from $r$ incident with $uv,vw,uw$ respectively.
It follows that $\beta_{e}(r),\gamma_{e}(r)\le 0$ for every edge $e$ of $r$. 
\\
\\
(1) {\em If $r_1$ is big then $\beta_{uv}(r) = -1$.}
\\
\\
For let us examine the $\beta$-rules. Certainly $uv$ is not a door for $r_1$, since $r$ is a triangle; so the first rule
does not apply. Let $f,f'$ be the edges incident with $r_1$ different from $uv$ that are incident with $u,v$ respectively.
If the second $\beta$-rule applies then $m(uv) = 2$ and $m(f), m(f') \ge 5$, which implies that $m(uw), m(vw) = 1$, 
contradicting that $uvw$ has multiplicity at least five. If the third rule applies, then $m(uv) = 2$
and $m^+(f) = 6$ and $m^+(f') = 5$ say; but then $m(uw) = 1$ and $m(vw) = 2$, contrary to Conf(1).
The fourth rule does not apply, by Conf(1). Thus we assume that the fifth rule applies. Let $m(uv) = 3$,
$m^+(f) = 5$, and $m^+(f')<5$. Hence $m(f) = 4$, and so $u$ has degree three, and $m(vw)= 1$ by Conf(2), and $r_3$ is small,
and $\beta_{uv}(r) = -1/2$. Since 
$$ m(uv)+m(vw)+m(uw)+\beta(r) + \gamma(r)>4$$
it follows that
$$\beta_{uw}(r)+\beta_{vw}(r)+\gamma_{uw}(r)+\gamma_{vw}(r)\ge 0,$$
and since all the terms on the left are non-positive it follows that they are all zero.
Now $r_2$ is not big since $\beta_{vw}(r) = 0$, and $r_3$ is not a triangle by Conf(2),
so the third $\gamma$-rule  applies to $uw$, a contradiction since $\gamma_{uw}(r) = 0$. This proves (1).

\bigskip

Let $X =\{u,v,w\}$. Since $(G,m)$ is prime, it follows that $|V(G)\setminus X|\ge 3$, and $m(\delta(X))\ge 10$.
But
$$m(\delta(X)) = m(\delta(u)) +m(\delta(v))+ m(\delta(w)) - 2 m(uv)-2m(uw)-2m(vw),$$
and so $10\le 8+8+8 - 2 m(uv)-2m(uw)-2m(vw)$, that is,
$r$ has multiplicity at most seven. 
Suppose first that $r$ has multiplicity seven. By Conf(3), none of $r_1,r_2,r_3$ is a triangle. Now $\beta(r) + \gamma(r) > -3$. 
Consequently we may assume that $\beta_{uv}(r)+\gamma_{uv}(r)>-1$, and hence $r_1$ is small by (1). By Conf(7),  $m(uv)+m(uw)< 6$ 
and hence $m(vw)\ge 2$; and similarly $m(uw)\ge 2$. Now $\gamma_{uv}(r)>-1/2$, and so the first, fourth and sixth $\gamma$-rules do not apply to $uv$.
Since the first $\gamma$-rule does not apply, $m(uv)>1$.
Since the sixth $\gamma$-rule does not apply, one of $m(uw),m(vw)>2$, say $m(uw)\ge 3$, and so 
$m(uv) = 2$, $m(uw) = 3$ and $m(vw) = 2$. Since the fourth $\gamma$-rule does not apply, $r_1$ has no door disjoint from $uv$, contrary to
Conf(8). 

Next, suppose that $r$ has multiplicity six. Thus $\beta(r) + \gamma(r) > -2$, and so by (1), at most one of $r_1,r_2,r_3$ is big.
Suppose that $m(uv) = 4$; then $m(vw),m(uw) = 1$. Since at most one of $r_1, r_2,r_3$ is big, it follows from Conf(7) that $r_1$ is big, and hence
$r_2,r_3$ are small. By Conf(3), $r_2,r_3$ are not tough. By the second $\gamma$-rule, $\gamma_{vw}(r) = \gamma_{uw}(r) = -1/2$,
and since $\beta_{uv}(r) = -1$ by (1), this contradicts $\beta(r) + \gamma(r) > -2$. Thus $m(uv)\le 3$. Suppose next that $m(uv) = 3$; then from the symmetry we may assume that
$m(uw) = 2$ and $m(vw) = 1$. Since one of $r_1,r_2$ is small, and $r_3$ is not tough by Conf(3), 
the first $\gamma$-rule implies that
$\beta_{vw}(r)+\gamma_{vw}(r)\le -1$. Since $\beta(r) + \gamma(r) > -2$, it follows from (1) that neither of $r_1,r_3$ is big, contrary to Conf(7).
Thus $m(uv)\le 2$, and similarly $m(uw), m(vw)\le 2$, and so $m(uv),m(uw), m(vw) = 2$. Since $\beta(r) + \gamma(r) > -2$, it follows that
$\beta_e(r)+\gamma_e(r)\le -1$ for at most one edge $e$ incident with $r$; and so we may assume that
$\beta_{uv}(r)+\gamma_{uv}(r)> -1$ and $\beta_{uw}(r)+\gamma_{uw}(r)> -1$. By (1), $r_1,r_3$ are both small. By Conf(3), $r_1,r_3$ are not tough, 
and since the fourth $\gamma$-rule does not apply, it follows that $r_1$ has at most one door, and no door disjoint from $uv$, and $r_3$ has at most one door, and no door disjoint
from $uw$, and $u$ has degree at least four, contrary to Conf(9).

Finally, suppose that $r$ has multiplicity five. Now $\beta(r) + \gamma(r) > -1$, and hence $\beta_e(r)+\gamma_e(r)>-1$ for every edge $e$ incident with $r$; and
so by (1) $r_1,r_2,r_3$ are all small. Suppose that $m(uv) = 3$, and hence $m(uw), m(vw) = 1$. If neither of $r_2,r_3$ is tough, then 
by the second $\gamma$-rule, $\gamma_{uw}(r) = \gamma_{vw}(r) = -1/2$, a contradiction. Thus we may assume that $r_3$ is a tough triangle $uwx$.
By Conf(5),  $m(wx)=1$, and so $m(ux) \ge 3$ since $r_3$ is tough, contrary to Conf(3).
Thus we may assume that $m(uv)\le 2$; and so from the symmetry we may assume that $m(uv) = m(uw) = 2$ and $m(vw) = 1$. 
The first $\gamma$-rule does not apply to $vw$, and so $r_2$ is a tough triangle $vwx$. By Conf(3), $m(vx), m(wx) \le 2$, and so $m(vx), m(wx) = 2$. 
Since $r_2$ is tough, one of $vx,wx$ is 
incident with a small region different from $uvx$, contrary to Conf(5). This proves \ref{triovercharge2}.~\bbox

\begin{thm}\label{smallovercharge}
If $r$ is a small region with length at least four and with $\alpha(r)+\beta(r)+\gamma(r)>0$, then $(G,m)$ is not prime.
\end{thm}
\Proof
Suppose that $(G,m)$ is prime. Let $C = C_r$.
Note that for each $e\in E(C)$, $-1\le \beta_e(r)\le 0$ and $0\le \gamma_e(r)\le 1$
Since $\alpha(r) = 8 - 4|E(C)| +  \sum_{e\in E(C)}m(e)$,
it follows that
$$8 - 4|E(C)| +  \sum_{e\in E(C)}m(e) + \sum_{e\in E(C)}(\beta_e(r)+\gamma_e(r))>0,$$
that is,
$$\sum_{e\in E(C)}(m(e)+\beta_e(r)+\gamma_e(r)-4) > -8.$$
For each $e\in E(C)$, let 
$$\phi(e) = m(e)+\beta_e(r)+\gamma_e(r).$$ 
It follows that $|\phi(e)-m(e)|\le 1$ for each $e$ by \ref{nonzero}. 
For each integer $i$, let $E_i$ be the set of edges of $C$ such that $\phi(e)\in \{i,i-\frac12\}$.
\\
\\
(1) {\em For every $e\in E(C)$, $\phi(e)$ is one of $0,\frac12,1,\frac32,2,\frac52,3,4$, and hence $E(C)$ is the union of $E_0,E_1,E_2,E_3,E_4$.}
\\
\\
For let $e\in E(C)$. Since  $m(e)\ge 1$ and $\beta_e(r)\ge -1$ it follows that $\phi(e)\ge 0$.
Next we show that $\phi(e) \le 4$.
Now $m(e)<6$ by Conf(14). Suppose that $m(e) = 5$. Then the second region incident with $e$
is big, by Conf(14); and hence $\beta_e(r)=-1$ from the $\beta$-rules, and $\gamma_e(r) = 0$
and so  $\phi(e) \le 4$. Now suppose that $m(e) = 4$. Then by the $\gamma$-rules,
$\gamma_e(r) = 0$, and so $\phi(e) \le 4$. Finally, if $m(e)\le 3$ then $\phi(e) \le 4$
since $\gamma_e(r)\le 1$. Thus $\phi(e)\le 4$ in all cases. Finally, suppose that $\phi(e) = \frac72$, and hence $m(e) = 3$ or $4$. If $m(e) = 3$
then $\gamma_e(r) = 1/2$, contrary to the $\gamma$-rules; while if $m(e) = 4$ then $\beta_e(r) = -1/2$, contrary to the $\beta$-rules. This proves (1).
\\
\\
(2) {\em Let $e\in E(C)$; then $e\in E_4$ if and only if either $m^+(e) \ge 5$, 
or $m(e) = 3$ and $e$ is $5$-heavy.
Moreover, no two edges in $E_4$ are consecutive in $C$.}
\\
\\
The first assertion is immediate from the $\beta$- and $\gamma$-rules. 
For the second, suppose that $e,f\in E_4$ share an end $v$. Since $v$ has degree at
least three, it follows that $m(e)+m(f)\le 7$ and so we may assume that $m(e) =3$. Let $e$ have ends $u,v$; then from the first assertion there is
a triangle $uvw$ where $m(uw), m(vw) = 2$. Hence $m(f)=3$, and so there is similarly a triangle containing $f$, with third vertex $x$. 
Consequently $w = x$; but this contradicts Conf(3) and hence proves (2).
\\
\\
(3) {\em If $e\in E_4$, and $f\in E(C)$ is disjoint from $e$, and 
every edge in $E(C)\setminus \{f\}$ disjoint from $e$ is $3$-heavy, and there is no edge of $C$ with multiplicity one disjoint from $f$, 
then $f\in E_0$.}
\\
\\
For by Conf(6) if $|E(C)| = 4$ and $m^+(e)\ge 5$, or by Conf(17) or Conf(18) otherwise, it follows that
$m^+(f) =1$. Since there is no edge of $C$ with multiplicity one disjoint from $f$,
it follows that $\beta_f(r) = -1$ from the $\beta$-rules, and so $f\in E_0$. This proves (3).

\bigskip
For $0\le i\le 4$, let $n_i = |E_i|$.
\\
\\
(4) {\em If $e\in E(C)$ satisfies $m(e) = 2$, and $n_4 = 0$, and $r$ has at most one door, and no door disjoint from $e$, then $\phi(e)\le 2$.}
\\
\\
For if not, then $\gamma_e(r)>0$, and so from the $\gamma$-rules, there is a triangle $uvw$ with $e = uv$, 
and some edge $f$ of $C$ consecutive with $e$ satisfies $m^+(f) = 5$;
but then $f\in E_4$, contradicting that $n_4 = 0$. This proves (4).
\\
\\
(5) {\em If $u,v,w$ are consecutive vertices in $C$, and $uv\in E_4$ and $m(uv) = 3$, then $\phi(vw)\le 2$.}
\\
\\
For since $uv\in E_4$, by (2) there is a triangle $uvx$ with $m(ux) = m(vx)= 2$. From Conf(2) it follows that
$m(vw)\le 2$; and since $w$ is not adjacent to $x$ by Conf(3), and hence $vw$ is not $4$-heavy, the $\gamma$-rules imply that $\phi(vw)\le 2$.
This proves (5).

\bigskip
Let $C$ have vertices $v_1\l v_k$ in order, and let $v_{k+1}$ mean $v_1$. For $1\le i\le k$ let $e_i$ be the edge $v_iv_{i+1}$, and let $r_i$ be the region
incident with $e_i$ different from $r$.

Since
$$\sum_{e\in E(C)}(\phi(e)-4) > -8,$$
we have
$4 n_0 +3n_1+2n_2+n_3 \le 7$, that is, 
$$3 n_0 +2n_1+n_2 + k  -n_4  \le 7,$$
since $n_0+n_1+n_2+n_3+n_4 = k$. But by (1), $n_4\le k/2 $ and so
$$3 n_0 +2n_1+n_2 + k/2  \le 7.$$
Since $k\ge 4$ it follows
that $3 n_0 +2n_1+n_2 \le 5$, and hence $n_0+n_1\le 2$.

\bigskip

\noindent{\bf Case 1:} $n_0+n_1=2$.

\bigskip

Since $3 n_0 +2n_1+n_2 + k  -n_4  \le 7,$  we have $n_4\ge n_0+n_2 +k-3$. Thus $n_4>0$.
If $k = 4$, let $e\in E_4$; then by (3) the edge $f$ of $C$ disjoint from $e$ belongs to $E_0$, and so by (2), $n_4 = 1$; but this contradicts $n_0+n_2 +k-3\le n_4$.

Thus $k\ge 5$. Since 
$$3 n_0 +2n_1+n_2 +  k/2  \le 7,$$
and $2n_0+2n_1 = 4$ and $ k/2\ge 5/2$, it follows that $n_0 = n_2 = 0$ and $n_1 = 2$ and $k \le 6$.

Suppose that $k = 6$; then $n_4 = 3$ since $n_4\ge n_0+n_2 +k-3$, so we may assume that $e_1,e_3,e_5\in E_4$.
By Conf(17) and Conf(18), it follows that $m^+(e_4) = 1$, 
and hence $e_4\in E_0\cup E_1$, and similarly $e_6,e_2\in  E_0\cup E_1$, a contradiction since $n_0+n_1=2$.
Thus $k = 5$, and so $n_4 \ge 2$, and by (2) $n_4 = 2$ and we may assume that $e_1,e_3\in E_4$. 
By Conf(17) and Conf(18), $m^+(e_4) = 1$, and similarly $m^+(e_5) = 1$. 
Since $n_1 = 2$, and $n_0,n_2 = 0$, it follows that $m(e_2)>1$.
But then $e_4\in E_0$ by (3), contradicting that $n_0 = 0$.

\bigskip

\noindent{\bf Case 2:} $k = 4$ and $n_0+n_1=1$ and $n_4>0$.

\bigskip

Let $e_4\in E_4$; by (3), $e_2\in E_0$ and so $m(e_2) = 1$.
By (2) and Conf(2) and Conf(4), it follows
that $m(e_1),m(e_3)\le 2$. Now $e_2$
is the only edge of $C$ that is not $2$-heavy, since $n_0+n_1=1$, and in particular $r$ has at most one door. 
Since $4 n_0 +3n_1+2n_2+n_3 \le 7$ and $n_0 = 1$, it follows that $n_2\le 1$, so we may assume that $e_1\notin E_2$. Thus $\phi(e_1)>2$, and 
hence $m(e_1) = 2$. By (2) and (5), 
$m^+(e_4)\ge 5$, so by Conf(4), $m(e_4) = 4$.
Since $\phi(e_1)>2$, it follows from the $\gamma$-rules that 
$r_1$ is  a triangle $v_1v_2w$ say, where $m(v_1w),m(v_2w)\ge 2$. Consequently $m(v_1w) = 2$. 
Since $e_3\notin E_1$, it follows that $m^+(e_3)\ge 2$; so $m(v_2w) = m^+(v_2w) = 2$ by Conf(18) 
(taking $v_2,v_1,w$ to be the vertices called $u,v,w$ in Conf(18) respectively). From Conf(10) it follows that $m(e_3) = 1$. From the $\gamma$-rules it follows that $\phi(e_1) = 5/2$.
Since $\sum_{e\in E(C)}\phi(e) > 8$ and $\phi(e_2)+\phi(e_4) \le 4$, it follows that $\phi(e_3)\ge 2$. Since $m(e_3) = 1$, the $\gamma$-rules imply that
$e_3$ is $3$-heavy, contrary to Conf(16) (taking $v_2,v_1,w$ to be the vertices called $u,v,w$ in Conf(16) respectively).
\bigskip

\noindent{\bf Case 3:} $k = 4$ and $n_0+n_1=1$ and $n_4=0$.

\bigskip

Let $e_4\in E_0\cup E_1$, and so $m(e_4)\le 2$. Since every edge of $C$ that is not $2$-heavy belongs to $E_0\cup E_1$,
it follows that $e_1,e_2,e_3$ are $2$-heavy. Since $n_4 = 0$, it follows that $m^+(e_i)\le 4$ for $i = 1,2,3,4$.

Suppose that $\phi(e_1)\ge 3$, and hence $\phi(e_1) = 3$ by (1) since $n_4 = 0$. By (4) it follows that $m(e_1)\ge 3$. If $m^+(e_1) = 3$, then from the $\beta$-rules,
the edge $xv_2$ of $r_1$ incident with $v_2$ and different from $e_1$ has multiplicity four and hence $m(e_2) = 1$; and since $x,v_3$ are non-adjacent
by Conf(2), this contradicts that $e_2$ is $2$-heavy. Thus $m^+(e_1)\ge 4$. By Conf(6), $m^+(e_3)\le 2$, and so $\phi(e_3)\le 2$ by (4). 
Since $\phi(e_2)\le 3$, and $\phi(e_4)\le 1$, and $\sum_{e\in E(C)}\phi(e) > 8$, it follows that $\phi(e_2)\ge 5/2$ (and so $e_2$ is $3$-heavy), 
and $\phi(e_3)\ge 3/2$, and $\phi(e_4)\ge 1/2$ (and so $m^+(e_4)\ge 2$). 
By Conf(2), it is not the case that $m(e_3) = 2$ and the edge of $r_3$ consecutive with $e_3$
and incident with $v_3$ has multiplicity four; and so, since $\phi(e_3)\ge 3/2$, the $\beta$-rules imply that $m(e_3) = 1$ and $r_3$ is a triangle $v_3v_4y$ say.
Now by Conf(15), not both $m(v_3y), m(v_4y)\ge 2$; and $m(e_2)\le 3$ by Conf(4), so by Conf(18), $m^+(v_3y), m^+(v_4y) \le 3$. But then the $\gamma$-rules imply that
$\phi(e_3)\le 1$, a contradiction.
This proves that $\phi(e_1)\le 5/2$; and similarly $\phi(e_3)\le 5/2$.

Since $\sum_{e\in E(C)}\phi(e) > 8$, and $\phi(e_2)\le 3$ (because $n_4 = 0$)
it follows that $\phi(e_1)+ \phi(e_3)\ge 9/2$, and $\phi(e_4)\ge 1/2$; and from the
symmetry we may assume that
$\phi(e_1) = 5/2$ and $\phi(e_3)\ge 2$.
The $\beta$- and $\gamma$-rules imply that $m(e_1) = 3$ (since $m^+(e_2)\le 4$). Since $\phi(e_2)+\phi(e_3) \ge 5$, and $\phi(e_3)\le 5/2$, it 
follows that $\phi(e_2)\ge 5/2$ (and hence $m(e_2)\ge 2$). 

Suppose that $m(e_3) = 1$. Since $\phi(e_3) \ge 2$, the first $\gamma$-rule applies, and so $r_3$ is a triangle $v_3v_4y$, and
$m(v_3y),m(v_4y)\ge 2$, and $m^+(v_3y)+m^+(v_4y)\ge 6$.
By Conf(4), $m(e_2)\le 3$, so by Conf(18), $m^+(v_3y),m^+(v_4y)\le 3$, and hence equality holds for both.  
By Conf(11), $m(v_3y),m(v_4y) = 2$; but this is contrary to Conf(16).

So $m(e_3)\ge 2$, and by Conf(4), $m(e_2) = m(e_3) = 2$. If $m^+(e_3) = 2$, then from the $\beta$-rules it follows that both edges of $r_3$ consecutive with $e_3$ have multiplicity five; but
this is impossible since $m(e_2)= 2$. So $m^+(e_3) = 3$.
Since $\phi(e_2)\ge 5/2$ it follows that
$r_2$ is a triangle $v_2v_3x$, $m(v_2x),m(v_3x)\ge 2$, and one of $m^+(v_2x), m^+(v_3x)\ge 3$, and $e_4$ is a door for $r$.
Since $\phi(e_4)>0$, we deduce that $m^+(e_4)\ge 2$.
By Conf(2), $m(v_2x) = 2$.
By Conf(12), $m^+(v_3x)=2$ and $m^+(v_2x) = 2$, a contradiction.

\bigskip

\noindent{\bf Case 4:} $k = 4$ and $n_0+n_1 = 0$.

\bigskip
Since $n_0,n_1 = 0$, it follows that $\phi(e_i)\ge 3/2$ and hence $e_i$ is $2$-heavy, for $1\le i\le 4$.
Consequently $n_4 = 0$, from (3). Since $\sum_{e\in E(C)}\phi(e) > 8$, we may assume because of the  
symmetries of the square that
$\phi(e_1)+\phi(e_3)\ge 9/2$, and $\phi(e_1)\ge\phi(e_3)$, and therefore
$\phi(e_1)\ge 5/2$. Thus $m(e_1)\ge 3$ from (4). 
If some edge $f$ of the boundary of $r_1$ consecutive with $e_1$ satisfies $m(f) = 4$,
say $f = v_1x$, then $m(e_4) = 1$ and $v_1$ has degree three; but since $e_4$ is $2$-heavy, it follows that $x,v_4$ are adjacent, contrary to Conf(2). 
Thus there is no such $f$, and so by the $\beta$-rules,
$m^+(e_1) \ge 4$.

Suppose that $m(e_3)\ge 2$. By Conf(6) it follows that $m^+(e_3) = 2$, and in particular $r_3$ is big.
Since $\phi(e_3)\ge 3/2$, the $\beta$-rules imply that some edge $f$ of the boundary of $r_3$ consecutive with $e_3$
satisfies $m(f) = 5$, say $f = v_4x$; and since $x,v_1$ are nonadjacent by Conf(2) it follows that $e_4\in E_0\cup E_1$,
a contradiction. Thus $m(e_3)=1$. Since $e_3$ is $2$-heavy it follows that $r_3$ is a triangle $v_3v_4x$ say.

By Conf(4), $m(e_2), m(e_4)\le 3$.
By Conf(15), we may assume that $m(v_3x) = 1$; and by Conf(18), $m^+(v_4x)\le 3$. Since $m(e_4)\le 3$, the $\gamma$-rules imply 
that $\phi(e_3)\le 1$, a contradiction.

\bigskip

\noindent{\bf Case 5:} $k\ge 5$ and $n_0+n_1 = 1$.

\bigskip

Since $3 n_0 +2n_1+n_2 + k  -n_4  \le 7,$ we have $n_4\ge n_0+n_2+k-5$.  Let $E_0\cup E_1 = \{e_k\}$.

Suppose that $n_4 = 0$. Then since $n_4\ge n_0+n_2+k-5$ it follows that $k = 5$. Since 
$$\sum_{e\in E(C)}\phi(e) > 4k-8 = 12,$$ and
$\phi(e_5)\le 1$, and $\phi(e_i)\le 3$ for $i = 1,2,3,4$ (by (1), since $n_4 = 0$)
it follows that $\phi(e_i)\ge 5/2$ for $i = 1,2,3,4$, and hence $e_1\l e_4$ are $3$-heavy. If $m(e_1)\le 2$, then since
$\phi(e_1)\ge 5/2$ it follows from the $\gamma$-rules that $m(e_2) = 4$ and $r_2$ is small; but then $e_2\in E_4$, a contradiction. 
Thus $m(e_1)\ge 3$; so $m(e_1) = m^+(e_1) = 3$
by Conf(15). Since $m(e_2)\ge 2$, it follows that not both edges of $r_1$ consecutive with $e_1$ have multiplicity four, and so from the $\beta$-rules,
$\phi(e_1) \le 5/2$. Similarly $\phi(e_4)\le 5/2$, contradicting that $\sum_{e\in E(C)}\phi(e) > 12$. This proves that $n_4>0$. 

Suppose that $n_2 = 0$. Thus $e_1\l e_4$ are $3$-heavy. Since $n_4>0$, (3) implies that $n_0 = 1$. Since $\phi(e_1)>2$, the $\beta$- and $\gamma$-rules imply that either:
\begin{itemize}
\item $m(e_1) = 2$ and $r_1$ is a triangle
$v_1v_2w$ say; and $m(v_1w),m(v_2w)\ge 2$, and $m(e_2) = 4$. Consequently $m(v_2w) = 2$, contrary to Conf(16). 
\item $m(e_1) = 3$ and $r_1$ is big, and, if  $u_1\d v_1\d v_2\d u_2$ is the three-edge path of $C_{r_1}$ with middle edge $e_1$, then
one of $m(u_1v_1),m(u_2v_2) = 4$ and is incident with a small region. But if $m(u_1v_1) = 4$ then the second region incident with it is $r_k$,
and this is not small since $n_0 = 1$; and if $m(u_2v_2) = 4$ then $v_2$ has degree three and $m(e_2) = 1$, and since $e_2$ is $3$-heavy it follows that $u_2,v_3$ are adjacent,
and $m(u_2v_3)\ge 2$, contrary to Conf(2).
\item $m^+(e_1)\ge 4$; but this is contrary to Conf(15).
\end{itemize}
This proves that $n_2 \ge 1$. 

Since
$3 n_0 +2n_1+n_2 + k/2  \le 7$,
we have $n_0+n_2 + k/2  \le 5,$ and in particular $n_2\le 2$.
If $e\in E(C)$ is not $3$-heavy, then $\phi(e)\le 2$ from the $\gamma$-rules, and so at most
two edges of $E(C)$ not in $E_0\cup E_1$ are not $3$-heavy.
By Conf(8) and Conf(19) it follows that $e_1,e_{k-1}\notin E_4$,
so every edge in $E_4$ is disjoint from $e_k$. 
Since there are three consecutive edges of $C$ not in $E_4$, and no two edges in $E_4$ are consecutive by (2), it follows that $n_4\le k/2-1$; and since
$3 n_0 +2n_1+n_2 + k  -n_4  \le 7$, it follows that $ n_0+n_2 + k/2   \le 4$, and so $n_2 = 1$, and $n_0 = 0$, and $k\le 6$.
In particular, from (5) every edge $e\in E_4$ has $m(e)\ge 4$.

Suppose that $k = 6$. Since $n_4\ge n_0+n_2+k-5$ and $n_4\le k/2-1$, it follows that $n_4 = 2$; and so $E_4 = \{e_2,e_4\}$, since the members of $E_4$ are disjoint
from $e_6$ and from each other. Since $e_2\in E_4$, (3) implies that $e_5$ is not $3$-heavy, and so $e_5\in E_2$; and similarly
$e_1\in E_2$, a contradiction since $n_2 = 1$.

Thus $k = 5$. Since $n_4\le k/2-1$ it follows that $n_4 = 1$, so we may assume that $E_4 = \{e_2\}$. By (3), $e_4$ is not $3$-heavy, and so $\phi(e_4)\le 2$.
Consequently $E_2 = \{e_4\}$, and $\phi(e_1)+\phi(e_3)\ge 11/2$. 
Since $\phi(e_4),\phi(e_5)>0$, it follows that $m^+(e_4), m^+(e_5)\ge 2$, and since $m^+(e_2)\ge 5$, 
two applications of Conf(13) imply that
$m(e_3)+m(e_4)\le 3$ and $m(e_1)+m(e_5)\le 3$. Since $m(e_1), m(e_3)\ge 2$ (because $\phi(e_1), \phi(e_3)>2$) it follows that $m(e_1), m(e_3) =2$ and $e_1,e_3$ are $4$-heavy; and 
$m(e_4), m(e_5) = 1$. Since $\phi(e_4)>1$, $r_4$ is a triangle $v_4v_5x$ say. Since $e_4$ is not $3$-heavy, one of $m(v_4x),m(v_5x) = 1$. If $m(v_4x) = 1$ then
by Conf(16), $m(xv_5)\le 2$; but then $\phi(e_4)= 1$ from the $\gamma$-rules, a contradiction. So $m(v_5x)= 1$. Since $\phi(e_4)>1$, the $\gamma$-rules imply that
$m^+(v_4x)\ge 4$. But this contradicts Conf(18).

\bigskip

\noindent{\bf Case 6:} $k\ge 5$ and $n_0+n_1 = 0$.

\bigskip

Since $n_0,n_1 = 0$, it follows that $\phi(e_i)\ge 3/2$ and hence $e_i$ is $2$-heavy, for $1\le i\le k$.
Since $3n_0 +2n_1+n_2 + k  -n_4  \le 7,$
we have
$n_4\ge n_2+k-7$.

Suppose first that $n_4>0$. By (2) and Conf(8) and Conf(19), every edge in $E_4$ is disjoint from at least three edges that are not
$3$-heavy and that therefore belong to $E_2$. In particular $n_2\ge 3$. 
Let $e\in E_4$; then $e$ is disjoint from all the other edges in $E_4$, and from at least three edges in $E_2$, so $k-3\ge n_4-1 + 3$, that is, $k\ge n_4+5$.
But $n_4\ge n_2+k-7\ge k-4$, a contradiction.

This proves that $n_4 = 0$, and so $E(C) = E_2\cup E_3$. Since $n_4\ge n_2+k-7$, it follows that $n_2+k\le 7$.
In particular, $k\in \{5,6,7\}$. From (4), every edge $e\in E(C)$ with $m(e) = 2$ belongs to $E_2$, since $n_4 = 0$ and
there are no doors for $r$. Consequently every $e\in E_3$ satisfies $m(e)\ge 3$.
Suppose that $m^+(e) = 3$ for some $e\in E_3$, say $e = e_1$. Thus $r_1$ is big, and $\beta_e(r)>-1$ since $\phi(e)>2$. Hence from the $\beta$-rules,
some edge of $C_{r_1}$ consecutive with $e_1$ has multiplicity four, say $v_1x$. Hence $m(e_k) = 1$, and since $n_0,n_1 = 0$, it follows that
$r_k$ is a triangle, and therefore $x,v_k$ are adjacent, contrary to Conf(2). This proves that $m^+(e)\ge 4$ for every $e\in E_3$.

By Conf(15), every edge in $E_3$ is disjoint from some edge in $E_2$, and in particular $n_2\ge 2$.
Since $n_2+k\le 7$, we have $k = 5$ and $n_2 = 2$. Every edge in $E_3$ is disjoint from one of the edges in $E_2$, so we may assume that
$e_1,e_2\in E_2$, and $e_3,e_4,e_5\in E_3$. Since $m^+(e_3),m^+(e_4), m^+(e_5)\ge 4$, Conf(13) implies that $m^+(e_1)\le 2$; and by Conf(15), $e_1$ is not $3$-heavy. 
From the $\gamma$-rules, $\phi(e_1)\le 3/2$, and similarly $\phi(e_2)\le 3/2$. But for $i = 3,4,5$, $\phi(e_i)\le 3$ since $n_4 = 0$; and so
$\sum_{e\in E(C)}\phi(e)\le 12$, contradicting our initial assumption that
$$\sum_{e\in E(C)}(\phi(e)-4) > -8.$$

\bigskip

This completes the proof of \ref{smallovercharge}.~\bbox

\noindent{\bf Proof of \ref{unav}.\ \ }
Suppose that $(G,m)$ is a prime $8$-target, and let $\alpha, \beta, \gamma$ be as before. Since the sum over all regions $r$ of
$\alpha(r)+\beta(r)+\gamma(r)$ is positive, there is a region $r$ with $\alpha(r)+\beta(r)+\gamma(r)>0$. But this is contrary to one of
\ref{bigovercharge}, \ref{triovercharge1}, \ref{triovercharge2}, \ref{smallovercharge}. This proves \ref{unav}.~\bbox

\section{Reducibility}

Now we begin the second half of the paper, devoted to proving the following.

\begin{thm}\label{reduc}
Every minimum $8$-counterexample is prime.
\end{thm}
Again, the proof is broken into several steps. Clearly no minimum $8$-counterexample $(G,m)$ has an edge $e$ with $m(e) = 0$, because deleting $e$
would give a smaller $8$-counterexample; and by \ref{counterex}, every minimum $8$-counterexample satisfies the conclusions of \ref{counterex}. Thus, it remains to check that 
$(G,m)$ contains none of Conf(1)--Conf(19). Sometimes it is just as easy to prove a result for general $d$ instead of $d = 8$, and so we do so.

\begin{thm}\label{touchtri}
If $(G,m)$ is a minimum $d$-counterexample, then every triangle has multiplicity less than $d$.
\end{thm}
\Proof
Let $uvw$ be a triangle of $G$, and let $X =\{u,v,w\}$. Since $|V(G)|\ge 6$, \ref{counterex} implies that $m(\delta(X))\ge d+2$.
But
$$m(\delta(X)) = m(\delta(u)) +m(\delta(v))+ m(\delta(w)) - 2 m(uv)-2m(uw)-2m(vw),$$
and so $d+2\le d+d+d - 2 m(uv)-2m(uw)-2m(vw)$, that is, $m(uv)+m(uw)+m(vw)\le d-1$.
This proves \ref{touchtri}.~\bbox

If $C$ is a cycle of length four in $G$, say with vertices $u,v,w,x$ in order, let $m'$ be defined as follows:
$m'(uv) = m(uv)-1$, $m'(vw) = m(vw)+1$, $m'(wx)= m(wx)-1$, $m'(ux) = m(ux)+1$, and $m'(e) = m(e)$ for all other edges $e$. If $(G,m)$ is a minimum $d$-counterexample, then
because
of the second statement of \ref{counterex}, it follows that $(G,m')$ is a $d$-target. (Note that possibly $m'(uv),m'(wx)$ are zero; this is the reason to
permit $m(e)=0$ in a $d$-target.) We say that $(G,m')$ is obtained from $(G,m)$ by {\em switching on the sequence $u\d v\d w\d x\d u$}.
If $(G,m')$ is smaller than $(G,m)$, we say that the sequence $u\d v\d w\d x\d u$ is {\em switchable}.

\begin{thm}\label{4conn}
No minimum $d$-counterexample contains Conf(1).
\end{thm}
\Proof
Suppose that $(G,m)$ is a minimum $d$-counterexample, with a triangle $uvw$, where $u,v$ have degree three. Let the neighbours of $u,v$ not in $\{u,v,w\}$ be $x,y$ respectively.
Let $H$ be a simple graph obtained from $G$ by adding new edges if necessary to make $w,x,y$ pairwise adjacent, and extend $m$ to $E(H)$
by setting $m(e) = 0$ for every new edge. Thus $(H,m)$ is not $d$-edge-colourable, and although it may not be a minimum $d$-counterexample,
no $d$-counterexample has fewer vertices. 

Define $f(w) = m(uw)+m(vw)$, $f(x) = m(ux)$, and $f(y) = m(vy)$. Since $m(\delta(\{u,v\}))$ is even, it follows that
$f(w)+f(x)+f(y)$ is even. Define 
\begin{eqnarray*}
n(wx) &=& \frac12(f(x)+f(w)-f(y))\\
n(wy) &=& \frac12(f(y)+f(w)-f(x))\\
n(xy) &=& \frac12(f(x)+f(y)-f(w)). 
\end{eqnarray*}
It follows that $n(wx),n(wy),n(xy)$ are integers. Since  $m(\delta(\{u,v,w\}))\ge d$ 
and $m(\delta(w)) = d$, it follows that $m(ux)+m(vy)\ge m(uw)+m(vw)$ and hence $n(xy)\ge 0$. Similarly, since
$m(\delta(\{u,v,x\}))\ge d$ and $m(\delta(x)) = d$, it follows that $n(wy)\ge 0$, and similarly $n(wx)\ge 0$.

Let $G' = H\setminus \{u,v\}$.
For each edge $e$ of $G'$, define $m'(e)$ as follows. If $e$ is incident with a vertex different from $x,y,w$ let $m'(e) = m(e)$.
For $e = xy,wx,wy$ let $m'(e) = m(e) + n(e)$.
We claim that $(G',m')$ is a $d$-target. To show this, let
$X\subseteq V(G')$ with $|X|$ odd; we must show that $m'(\delta_{G'}(X))\ge d$. By replacing $X$ by its complement if necessary (which also
is odd, since $|V(G)|$ is even), we may assume that $X$ contains at most one of $w,x,y$. But then from the choice of $f(w), f(x), f(y)$, 
it follows that $m'(\delta_{G'}(X)) = m(\delta_G(X))\ge d$ as required. Thus $(G',m')$ is a $d$-target. Since $|V(G')|<|V(G)|$, there are $d$
perfect matchings $F_1'\l F_d'$ of $G'$ such that every edge $e\in E(G')$ is in exactly $m'(e)$ of them. Now each of $F_1'\l F_d'$
contains at most one of the edges $wx,wy,xy$. Let $I_1,I_2,I_3, I_0$ be the sets of $i\in \{1\l d\}$ such that $F_i'$
contains $wx,wy,xy$ or none of the three, respectively. Thus 
$|I_1| = m'(wx) =m(wx)+n(wx) $. For $n(wx)$ values of $i\in I_1$ let $F_i = (F_i'\setminus\{wx\})\cup \{ux,vw\}$, and for the remaining $m(wx)$
values let $F_i = F_i'\cup \{uv\}$. Thus $F_i$ is a perfect matching of $G$ for each $i\in I_1$. Define $F_i\;(i\in I_2)$ similarly. For $n(xy)$
values of $i\in I_3$ let $F_i= ( F_i'\setminus \{xy\})\cup \{ux,vy\}$, and for the others let $F_i = F_i\cup \{uv\}$. For $i\in I_0$
let $F_i = F_i'\cup \{uv\}$. Then $F_1\l F_d$ are perfect matchings of $G$, and we claim that every edge $e$ is in exactly $m(e)$ of them.
This is clear if $e$ has an end different from  $u,v,w,x,y$; and true from the construction if both ends of $e$ are in $\{w,x,y\}$. From the symmetry
we may therefore assume that $e$ is incident with $u$. If $e = ux$, then $e$ belongs to 
$n(wx)+n(xy)$ of $F_1\l F_d$; but 
$$n(wx)+n(xy) = \frac12(f(x)+f(w)-f(y)) + \frac12(f(x)+f(y)-f(w)) = f(x) = m(ux)$$
as required. The other two cases are similar. This is a contradiction, since $(G,m)$ is a minimum $d$-counterexample, and so there is no such triangle $uvw$.
This proves \ref{4conn}.~\bbox

Incidentally, a similar proof would show that $G$ is four-connected except for cutsets of size three that cut off just one vertex, but we do not need this.

If $(G,m)$ is a $d$-target, and $x,y$ are distinct vertices both incident with some common region $r$, we define $(G,m)+xy$ to be
the $d$-target $(G',m')$ obtained as follows:
\begin{itemize}
\item If $x,y$ are adjacent in $G$, let $(G',m') = (G,m)$.
\item If $x,y$ are non-adjacent in $G$, let $G'$ be obtained from $G$ by adding a new edge $xy$, extending the drawing of $G$ to one of $G'$ and setting $m'(e) = m(e)$
for every $e\in E(G)$ and $m'(xy) = 0$.
\end{itemize}

\begin{thm}\label{cubic}
No minimum $d$-counterexample contains Conf(2).
\end{thm}
\Proof
Let $(G,m)$ be a minimum $d$-counterexample, with a triangle $uvw$, and suppose that 
$u$ has only one other neighbour $x$, and $m(ux)<m(uw)+m(vw)$. Let $(G',m'') = ((G,m) + vx)+wx$.
For each $e\in E(G')$,
define $m'(e)$ as follows. If $e\ne ux,uw,vw,vx$ let $m'(e) = m(e)$. Let 
\begin{eqnarray*}
m'(vx) &=& m''(vx)+m(vw)\\
 m'(vw)&=& 0\\
m'(ux) &=& m(ux)-m(vw)\\
m'(uw) &=& m(uw)+m(vw).
\end{eqnarray*}
Since $m(uv)+m(uw)+m(ux) = d$ and $m(uv)+m(uw)+m(vw)\le d$ since $m(\delta(\{u,v,w\}))\ge d$, it follows that $m(ux)\ge m(vw)$, and so $m'(e)\ge 0$
for every edge $e$. Moreover, $m'(\delta(z)) = d$ for every vertex $z$, from the construction. We claim that $(G',m')$ is a $d$-target. For let
$X\subseteq V(G')$ with $|X|$ odd; and we may assume that $u\notin X$. We must show that $m'(\delta(X))\ge d$. If $X$ contains at most one
of $v,w,x$ then $m'(\delta(X)) =m(\delta(X))\ge d$
as required, so we may assume that $X$ contains at least two of $v,w,x$. If $v,w,x\in X$ then $m'(\delta(X))\ge m'(\delta(u)) = d$ as required.
If $X\cap \{v,w,x\} = \{v,w\}$ then $m'(\delta(X)) =m(\delta(X)) +2m(vw)\ge d$, and if $X\cap \{v,w,x\} = \{w,x\}$ then $m'(\delta(X)) =m(\delta(X))\ge d$,
so we may assume that $X\cap \{v,w,x\} = \{v,x\}$, and hence $m'(\delta(X)) =m(\delta(X)) -2m(vw)$. We must therefore show that in this case,
$m(\delta(X))\ge 2m(vw)+d$. To see this, note that 
\begin{eqnarray*}
m(\delta(X\cup \{u,w\}))&=& m(\delta(X)) - m(ux)-m(uv)-m(vw)-m''(xw)\\
&& +(d-m(uw)-m(vw)-m''(xw)) \le m(\delta(X)) -2m(vw)
\end{eqnarray*}
since $m''(xw)\ge 0$ and $m(ux)+ m(uv) + m(uw) = d$.
Since $m(\delta(X\cup \{u,w\}))\ge d$, it follows that $m(\delta(X))\ge 2m(vw)+d$ as required.
This proves that $(G',m')$ is a $d$-target. Since $m'(uw)>m(ux),m(vw)$ (the first from the hypothesis), it follows that $(G',m')$ is smaller than $(G,m)$,
and so is $d$-edge-colourable; let $F_1'\l F_d'$ be a $d$-edge-colouring.
Now every perfect matching containing $vx$ also contains $uw$, since $vx$ is not disjoint from any other
edge incident with $u$. Hence there are at least $m(vw)$ of $F_1'\l F_d'$ that contain both $vx$ and $uw$. Choose $m(vw)$ of them, say $F_1'\l F_{m(vw)}'$; and
for $1\le i \le m(vw)$ define $F_i = (F_i'\setminus\{vx,uw\})\cup \{vw,ux\}$. Define $F_i = F_i'$ for $m(vw)+1\le i\le d$. Then
every edge $e$ of $G$ is in $m(e)$ of $F_1\l F_d$, a contradiction. Thus there is no such triangle $uvw$. This proves \ref{cubic}.~\bbox

\begin{thm}\label{dense}
No minimum $8$-counterexample contains Conf(3) or Conf(4).
\end{thm}
\Proof
To handle both cases at once, let us assume that $(G,m)$ is an $8$-target, and $uvw,uwx$ are triangles with $m(uv) + m(uw) + m(vw)+ m(ux)\ge 8$,
(where possibly $m(uw) = 0$); and either $(G,m)$ is a minimum $8$-counterexample, or $m(uw) = 0$ and deleting $uw$ gives a minimum $8$-counterexample $(G_0,m_0)$ say.
We must show that $m(uw) = 0$ and $(m(uv),m(vw),m(wx),m(ux))= (4,2,1,2)$.
Let $(G,m')$ be obtained by switching $(G,m)$ on $u\d v\d w\d x\d u$.
\\
\\
(1) {\em $(G,m')$ is not smaller than $(G,m)$.}
\\
\\
Because suppose it is. Then it admits an $8$-edge-colouring; because if $(G,m)$ is a minimum $8$-counterexample this is clear, and otherwise $m(uw) = 0$, and
$(G',m')$ is smaller than $(G_0,m_0)$. Let $F_1'\l F_8'$ be an $8$-edge-colouring of $(G',m')$.
Since 
$$m'(uv)+m'(uw)+m'(vw)+m'(ux)\ge 9,$$
one of $F_1'\l F_8'$, say $F_1'$, contains two of $uv,uw,vw,ux$ and hence contains $vw,ux$. Then 
$$(F_1'\setminus\{vw,ux\})\cup \{uv,wx\}$$
is a perfect matching, and it together with $F_2'\l F_8'$ provide an $8$-edge-colouring of $(G,m)$, a contradiction. This proves (1).

\bigskip

From (1) we deduce that $\max(m(ux),m(vw))<\max(m(uv),m(wx))$. It follows that
$$m(uv) + m(uw) + m(vw)+ m(wx)\le 7,$$ by (1) applied with $u,w$ exchanged;
and 
$$m(uv)+ m(ux)+m(wx) + m(uw)\le 7,$$ by (1) applied with $v,x$ exchanged. Consequently $m(ux)>m(wx)$, and hence $m(ux)\ge 2$; and
$m(vw)>m(wx)$, and so $m(vw)\ge 2$. Suppose that $m(uv)\le 3$.
Since 
$$\max(m(ux),m(vw))<\max(m(uv),m(wx)),$$ 
it follows that $m(uv) = 3$ and $m(vw)=m(ux) = 2$; and therefore $m(wx) = 1$, since $m(ux)>m(wx)$.
But this is contrary to (1).

We deduce that $m(uv)\ge 4$. Since $m(vw)\ge 2$ and $m(uv) + m(uw) + m(vw)+ m(wx)\le 7$,
it follows that $m(uw)+m(wx)\le 1$; so $m(uw) = 0$ and $m(wx) = 1$. But then 
$$(m(uv),m(vw),m(wx),m(ux))= (4,2,1,2).$$ 
This proves \ref{dense}.~\bbox

\section{Guenin's cuts}

We still have many configurations to handle, to finish the proof of \ref{reduc}, but all the others are handled by a method of Guenin~\cite{guenin},
which we introduce in this section. In particular, nothing so far has assumed the truth of \ref{mainconj} for $d = 7$, but now we will need to use that.

Let $(G,m)$ be a $d$-target, and let $x\d u\d v\d y$ be a three-edge path of $G$, where $x,y$ are incident with a common region. 
Let $(G',m')$ be obtained from $(G,m)+xy$ by switching on the cycle $x\d u\d v\d y\d x$.
We say that $(G',m')$ is obtained from $(G,m)$ by {\em switching on $x\d u\d v\d y$}.
If $(G',m')$ is smaller than $(G,m)$, we say that the path $x\d u\d v\d y$ is {\em switchable}.

Let $G$ be a three-connected graph drawn in the plane, and let $G^*$ be its dual graph; let us identify $E(G^*)$ with $E(G)$ in the natural way.
A {\em cocycle} means the edge-set of a cycle of the dual graph; thus, $Q\subseteq E(G)$ is a cocycle of $G$
if and only if $Q$ can be numbered $\{e_1\l e_k\}$ for some $k\ge 3$ and there are distinct regions $r_1\l r_k$ of $G$ such that
$1\le i\le k$, $e_i$ is incident
with $r_i$ and with $r_{i+1}$ (where $r_{k+1}$ means $r_1$). 

Guenin's method is the use of the following:

\begin{thm}\label{cuts}
Suppose that $d\ge 1$ is an integer such that every $(d-1)$-regular oddly $(d-1)$-edge-connected planar graph is $(d-1)$-edge-colourable.
Let $(G,m)$ be a minimum $d$-counterexample, and let $x\d u\d v\d y$ be a path of $G$ with $x,y$ on a common region. 
Let $(G',m')$ be obtained by switching on $x\d u\d v\d y$, and
let $F_1\l F_d$ be a $d$-edge-colouring of $(G',m')$, where $xy\in F_k$. 
Let $I = \{1\l d\}\setminus \{k\}$ if $xy\notin E(G)$, and $I = \{1\l d\}$ if $xy\in E(G)$.
Then for each $i\in I$,  there is a cocycle $Q_i$ of $G'$ with the following properties:
\begin{itemize}
\item for $1\le j\le d$ with $j \ne i$, $|F_j\cap Q_i| = 1$;
\item $|F_i\cap Q_i|\ge 5$;
\item there is a set $X\subseteq V(G)$ with $|X|$ odd such that $\delta_{G'}(X) = Q_i$; and
\item $uv,xy\in Q_i$ and $ux,vy\notin Q_i$.
\end{itemize}
\end{thm}
\Proof
Let $i\in I$. If $i\ne k$ and $xy\in F_i$, it follows that $m'(xy)\ge 2$ since $xy\in F_k$; and so $xy\in E(G)$.
Thus in either case $F_i$ is a perfect matching of $G$. For each edge $e$ of $G'$, let $p(e) = 1$ if $e\in F_i$, and $p(e) = 0$ otherwise; and
for each edge $e$ of $G$, let $n(e) = m(e)-p(e)$. Thus $(G,n)$
has the property that for each vertex $z$, $n(\delta_{G}(z)) = d-1$. If there is a list of $d-1$ perfect matchings of $G$ such that every edge $e$ is in $n(e)$
of them, then adding $F_i$ to this list gives a $d$-edge-colouring of $(G,m)$, a contradiction. Thus by hypothesis, there
exists $Y\subseteq V(G)$ with $|Y|$ odd and with $n(\delta_G(Y))<d-1$. Since $|Y|$ and $n(\delta_G(Y))$ have the same parity, it follows
that  $n(\delta_G(Y))\le d-3$. Since $\delta_G(Y)$ is an edge-cut of the connected graph $G$, it can be partitioned into ``bonds'' (edge-cuts $\delta_G(X)$
such that $G|X$, $G\setminus X$ are both connected), and hence one of these bonds $\delta_G(X)$ has $n(\delta_G(X))$ odd, and consequently $|X|$ also odd.
Since $\delta_G(X)$ is a bond of $G$ and hence $\delta_{G'}(X)$ is a bond of $G'$, there is a cocycle $Q_i$ of $G'$ with $Q_i = \delta_{G'}(X)$.
We claim that $Q_i$ satisfies the theorem. For we have seen the third assertion; we must check the other three. 

From the choice of $X$ we have $n(\delta_G(X)) \le d-3$. Since $|X|,|V(G)\setminus X|\ge 3$ (because $n(\delta_{G}(z)) = d-1$ for each vertex $z$),
it follows from \ref{counterex} that $m(\delta_G(X))\ge d+2$, and so  $p(\delta_G(X))\ge 5$, that is, $|F_i\cap Q_i|\ge 5$. This proves the second assertion. 
We recall that $F_1\l F_d$ is a $d$-edge-colouring of $(G', m')$; and so for $1\le j\le d$ with $j\ne i$, some edge of $\delta_{G'}(X)$ belongs to $F_j$, and
so 
$$\sum_{1\le j\le d, j\ne i}|F_j\cap Q_i|\ge d-1.$$
On the other hand, every edge $e$ of $G'$ belongs to $m'(e)$ of $F_1\l F_d$, and hence to $m'(e)-p(e)$ of the $d-1$ perfect matchings in this list without $F_i$.
Consequently
$$\sum_{1\le j\le d, j\ne i}|F_j\cap Q_i| = \sum_{e\in Q_i}m'(e)-p(e).$$
It follows that
$\sum_{e\in Q_i}m'(e)-p(e)\ge d-1;$
but $m'(e)-p(e) = n(e)$ for all edges of $G'$ except $xu,uv,vy,xy$, and so 
$$ |\{uv,xy\}\cap Q_i|-|\{ux,vy\}\cap Q_i| + \sum_{e\in Q_i}n(e) \ge d-1.$$
Since $\sum_{e\in Q_i}n(e) \le d-3$, it follows that $uv,xy\in Q_i$ and $ux,vy\notin Q_i$. This proves the fourth assertion. Moreover, since
$$\sum_{1\le j\le d, j\ne i}|F_j\cap Q_i|= d-1,$$
it follows that $|F_j\cap Q_i| = 1$ for all $j\in \{1\l d\}$ with $j\ne i$. This proves the first assertion, and so proves \ref{cuts}.~\bbox

\bigskip
By the result of~\cite{kawa}, every $7$-regular oddly $7$-edge-connected planar graph is $7$-edge-colourable, so we can apply \ref{cuts} when $d = 8$.

\begin{thm}\label{squareopp}
No minimum $8$-counterexample contains Conf(5) or Conf(6).
\end{thm}
\Proof
To handle both at once, let us assume that $(G,m)$ is an $8$-target, and $uvw,uwx$ are two triangles with $m^+(uv)+m(uw)+m^+(wx)\ge 7$; and either $(G,m)$ is a 
minimum $8$-counterexample, or
$m(uw) = 0$ and deleting $uw$ gives a minimum $8$-counterexample. 
We claim that  $u\d x\d w\d v\d u$ is switchable. For suppose not; then we may assume that $m(vw)>\max(m(uv),m(wx))$ and $m(vw)\ge m(ux)$. Now 
since one of $m(uv),m(wx)\ge 3$, and \ref{dense} implies that we do not have Conf(3) or Conf(4), it follows that
$$m(uv)+m(uw)+m(vw)+m(wx)\le 7.$$
Yet $m(uv)+m(uw)+m(wx) \ge 5$ since $m^+(uv)+m(uw)+m^+(wx)\ge 7$; and so $m(vw)\le 2$. Consequently $m(uv),m(wx) = 1$, and $m(ux)\le 2$. Since
$u\d x\d w\d v\d u$ is not switchable, it follows that $m(ux) = 2$; and since $m^+(uv)+m(uw)+m^+(wx)\ge 7$, it follows that $m(uw)\ge 3$, giving Conf(3), contrary to \ref{dense}.
This proves that  $u\d x\d w\d v\d u$ is switchable.

Let $r_1,r_2$ be the second regions  incident with $uv,wx$ respectively, and for $i = 1,2$ let $D_i$ be the set of doors for $r_i$.
Let $k = m(uv)+m(uw)+m(wx)+2$. Let $(G,m')$ be obtained by switching, and let $F_1\l F_8$ be an $8$-edge-colouring of $(G,m')$, where $F_i$ contains one of
$uv,uw,wx$ for $1\le i\le k$. For $1\le i\le 8$, let $Q_i$ be as in \ref{cuts}. 
\\
\\
(1) {\em For $1\le i\le 8$, either $F_i\cap Q_i\cap D_1\ne \emptyset$, or $F_i\cap Q_i\cap D_2\ne \emptyset$; and both are nonempty if either $k = 8$ or $i=8$.}
\\
\\
For let the edges of $Q_i$ in order be $e_1\l e_n,e_1$,
where $e_1 = wx$, $e_2 = uw$, and $e_3 =uv$. Since
$F_j$ contains one of $e_1,e_2,e_3$ for $1\le j\le k$, it follows that none of $e_4\l e_n$ belongs to any $F_j$ with $j\le k$ and $j\ne i$, and, 
if $k=7$ and $i\ne 8$, that only one of them is in $F_8$. But since at most one of $e_1,e_2,e_3$ is in $F_i$ and $|F_i\cap Q_i|\ge 5$, it follows that $n\ge 7$; so
either $e_4,e_5$ belong only to $F_i$, or $e_n, e_{n-1}$ belong only to $F_i$, and both if $k = 8$ or $i = 8$. But if $e_4,e_5$ are only contained in $F_i$, then
they both have multiplicity one, and are disjoint, so $e_4$ is a door for $r_1$ and hence $e_4\in F_i\cap Q_i\cap D_1$. Similarly if $e_n,e_{n-1}$ are only
contained in $F_i$ then $e_n\in F_i\cap Q_i\cap D_2$. This proves (1).

\bigskip

Now $k\le 8$, so one of $r_1,r_2$ is small since $m^+(uv)+m(uw)+m^+(wx)\ge 7$; and if $k = 8$ then by (1) $|D_1|,|D_2|\ge 8$, a contradiction. Thus $k = 7$, 
so both $r_1,r_2$ are small, but from (1) $|D_1|+|D_2|\ge 9$, again a contradiction. 
This proves \ref{squareopp}.~\bbox

\begin{thm}\label{bigtri}
No minimum $8$-counterexample contains Conf(7).
\end{thm}
\Proof
Let $(G,m)$ be a minimum $8$-counterexample, and suppose that $uvw$ is a triangle
with $m^+(uv)+m^+(uw) \ge 7$. Let $r_1,r_2$ be the second regions for $uv,uw$ respectively, and for $i = 1,2$
let $D_i$ be the set of doors for $r_i$. By \ref{squareopp}, we do not have Conf(5), so neither of $r_1,r_2$ is a triangle. Since $m(uv)+m(uw)\ge 5$, one of
$m(uv), m(uw)\ge 3$, so we may assume that $m(uv)\ge 3$. Let $tu$ be the edge incident with $r_2$ different from $uw$. Since  $m(uv)+m(uw)\ge 5$, it follows that
$m(tu)\le 3$, and by \ref{touchtri}, $m(vw)\le 2$. Thus the path $t\d u\d v\d w$ is switchable. Note that $t,w$ are non-adjacent in $G$, since $r_2$
is not a triangle.
Let $(G',m')$ be obtained by switching on this path, and
let $F_1\l F_8$ be an $8$-edge-colouring of it. Let $k = m(uv)+m(uw)+2$; thus $k\ge 7$, since $m(uv)+m(uw)\ge 5$, and we may assume that for $1\le j<k$, $F_j$ contains one of
$uv,uw$, and $tw\in F_k$. 

Let $I = \{1\l 8\}\setminus\{k\}$, and for each $i\in I$, let $Q_i$ be as in \ref{cuts}. Now let $i\in I$, and let the edges of $Q_i$ in order be $e_1\l e_n,e_1$, where $e_1 = uv$,
$e_2 = uw$, and $e_3 = tw$. Since $F_j$ contains one of $e_1,e_2,e_3$ for $1\le j\le k$ it follows that none of $e_4\l e_n$ belong to any $F_j$ with $j\le k$; and if
$k = 7$ and $i\ne 8$, only one of them belongs to $F_8$. Since $F_i$ contains at most one of $e_1,e_2,e_3$ and $|F_i\cap Q_i|\ge 5$, it follows that $n\ge 7$, and so
either $e_4,e_5$ are only contained in $F_i$, or $e_n, e_{n-1}$ are only contained in $F_i$; and both if either $k = 8$ or $i = 8$. 
Thus either $e_4\in F_i\cap Q_i\cap D_2$ or $e_n\in F_i\cap Q_i\cap D_1$, and both if $k = 8$ or $i = 8$. Since $k\le 8$, 
one of $r_1,r_2$ is small since $m^+(uv)+m^+(uw) \ge 7$; and yet if $k = 8$ then $|D_1|,|D_2| \ge  |I| =  7$, a contradiction. Thus $k=7$, so $r_1,r_2$ are both small, and yet
$|D_1|+|D_2|\ge 8$, a contradiction.
This proves \ref{bigtri}.~\bbox

\begin{thm}\label{farstep}
No minimum $8$-counterexample contains Conf(8).
\end{thm}
\Proof
Let $(G,m)$ be a minimum $8$-counterexample, and suppose that $uvw$ is a triangle, and its edges have multiplicities $3,2,2$ (in some order).
We will show that the second region $r$ for $uw$ has a door disjoint from $uw$.
By \ref{dense}, we do not have Conf(3), so $r$ is not a triangle. By exchanging $u,w$ if necessary we may assume that $m(vw) = 2$.
Let $tu$ be the edge incident with $r$ different from $uw$. We claim that the path $t\d u\d v\d w$ is switchable. For certainly $m(uv)\ge m(vw)$, so
it suffices to check that $m(uv)\ge m(tu)$. If not, then since $m(uv)\ge 2$ and $m(uv)+m(uw)\ge 5$, it follows that $m(uv) = 2$, $m(tu) = 3$
and $m(uw) = 3$, and we have Conf(2), contrary to \ref{cubic}. Thus $t\d u\d v\d w$ is switchable.
Let $(G',m')$ be obtained by switching, and let $F_1\l F_8$ be an $8$-edge-colouring of $(G',m')$. 
Since $m'(uv)+m'(uw) = 6$, we may assume that $F_1\l F_6$ each contain one of $uv,uw$; and $tw\in F_7$, and therefore $vw\in F_8$.
Let $I = \{1\l 6,8\}$; and for $i\in I$, let $Q_i$ be as in \ref{cuts}. Since $Q_8$ contains $uv,uw,tw$ and 
$F_1\l F_7$ each contain one of $uv,uw,tw$, it follows that no other edge of $Q_8$ belongs to any of $F_1\l F_7$, and so $Q_8\cap F_8$
contains a door for $r$, say $e$. Moreover $e\ne tu$ since $tu\notin Q_8$; and $e$ is not incident with $w$ since $vw\in F_8$. Consequently
$e$ is disjoint from $uw$. This proves \ref{farstep}.~\bbox

\begin{thm}\label{twofarstep}
No minimum $8$-counterexample contains Conf(9).
\end{thm}
\Proof
Let $(G,m)$ be a minimum $8$-counterexample, and suppose that $uv_1v_2$ is a triangle,
with $m(uv_1), m(uv_2), m(v_1v_2) = 2$, such that the second regions $r_1,r_2$ for $uv_1,uv_2$ respectively
both have at most one door, and no door that is disjoint from $uv_1v_2$. For $i = 1,2,$ let $D_i$ be the set of doors for $r_i$. For
$i = 1,2$, let $ux_i$ and $v_iy_i$ be edges incident with $r_i$ different from $uv_i$. 

Now $x_1\ne x_2$ since $u$ has degree at least four; and so 
$m(ux_1)+m(ux_2)\le 4$ and we may assume that $m(ux_1)\le 2$. Consequently the path $x_1\d u\d v_2\d v_1$ is switchable. 
Note that $v_1,x_1$ may be adjacent, but if so then $m(v_1x_1)= 1$ from \ref{touchtri}.
Let $(G',m')$ be obtained by switching, and let $F_1\l F_8$ be an $8$-edge-colouring, where
$uv_2\in F_1,F_2,F_3$, and $uv_1\in F_4,F_5$ and $v_1x_1\in F_6$, and $v_1x_1\in F_7$ if $v_1x_1\in E(G)$. 
Since $v_1v_2$ belongs to some $F_i$, and $v_1v_2$ meets all of
$uv_2,uv_1,v_1x_1$, we may assume that $v_1v_2\in F_8$. Let $I = \{1\l 5,7,8\}$ if $x_1v_1\notin E(G)$, and $I = \{1\l 8\}$ otherwise.
For $i\in I$, let $Q_i$ be as in \ref{cuts}. 

We claim that $F_i\cap Q_i\cap (D_1\cup D_2)\ne \emptyset$ for $i = 7,8$.
First suppose that $v_1x_1\notin E(G)$. 
Then for $1\le j\le 6$ and for $i = 7,8$, $F_j\cap Q_i\cap \{uv_2,uv_1,v_1x_1\}\ne \emptyset$,
and so no other edges of $Q_i$ belong to any $F_j$ with $j\in \{1\l 6\}$. Since only one edge of $Q_i\setminus \{uv_2,uv_1,v_1x_1\}$ belongs to the  $F_j$ with $j\in \{7,8\}\setminus \{i\}$,
it follows that $F_i\cap Q_i\cap (D_1\cup D_2)\ne \emptyset$ as required. Now suppose that $v_1x_1\in E(G)$. Then for 
 $1\le j\le 7$ and for $i = 7,8$, $F_j\cap Q_i\cap \{uv_2,uv_1,v_1x_1\}\ne \emptyset$. 
and so no other edges of $Q_i$ belong to any $F_j$ with $j\in \{1\l 7\}$ and $j\ne i$. For $i = 7$, as before it follows that
$F_i\cap Q_i\cap (D_1\cup D_2)\ne \emptyset$; for $i = 8$ we find that $F_i\cap Q_i\cap D_1 ,F_i\cap Q_i\cap D_2\ne \emptyset$.
Thus in any case, we have $F_i\cap Q_i\cap (D_1\cup D_2)\ne \emptyset$ for $j = 7,8$. 

Now by hypothesis, 
$D_1\cup D_2\subseteq \{ux_1,ux_2, v_1y_1,v_2y_2\}$; and $ux_1\notin Q_7, Q_8$ from the choice of switchable path, and
$v_1y_1,v_2y_2\notin F_8$ since $v_1v_2\in F_8$. Thus $ux_2\in F_8\cap D_2$. Since $|D_2|\le 1$ by hypothesis, it follows that $v_2y_2\notin D_2$,
and $ux_2\notin F_7$ since $ux_2\in F_8$ and $m(ux_2) = 1$. Thus $v_1y_1\in D_1$. Now $m(ux_2) = 1$, and so the path $x_2\d u\d v_1\d v_2$
is switchable; so by the same argument with $v_1,v_2$ exchanged, it follows that $ux_1\in D_1$ and $v_2y_2\in D_2$, contrary to the hypothesis.
This proves \ref{twofarstep}.

\begin{thm}\label{conf10}
No minimum $8$-counterexample contains Conf(10).
\end{thm}
\Proof
For suppose that $(G,m)$ is a minimum counterexample, with a square $uvwx$ and a triangle $wxy$, where $m(uv) = m(wx)=m(xy)=2$, and $m(vw) = 4$.
By \ref{dense}, we do not have Conf(4), and it follows that $m(ux) = 1$. Since $m(\delta(w)) = 8$ it follows that $m(wy)\le 2$, and so $u\d x\d y\d w$ is switchable.
Let $(G',m')$ be obtained by switching on this path, and let $F_1\l F_8$ be an $8$-edge-colouring of it. We may assume that
$xy\in F_1,F_2,F_3$, and $xw\in F_4,F_5$, and $uw\in F_6$. Let $I = \{1\l 8\}\setminus\{6\}$, and let $Q_i\;(i\in I)$ be as in \ref{cuts}.
Now $vw\notin F_4,F_5,F_6$, so there are four values of $i\in \{1,2,3,7,8\}$ such that $vw\in F_i$, and from the symmetry we may assume that
$F_1,F_2,F_7$ contain $vw$ (and so does one of $F_3,F_8$). It follows that $vw\notin Q_i$ for $i\in I$, and so $uv\in Q_i$ for each $i\in I$.
Since $uv$ belongs to two of $F_1\l F_8$, there exists $j\ne 8$ with $uv\in F_j$. Moreover, $F_j$ does not contain $vw$, and so $j\ne 1,2,7$;
so $j\in \{3,4,5,6\}$. But $|Q_1\cap F_j|\ge 2$, since one of $xy,xw,vw\in Q_1\cap F_j$, a contradiction. This proves \ref{conf10}.~\bbox

\begin{thm}\label{5gon}
No minimum $8$-counterexample contains Conf(11), Conf(12) or Conf(13).
\end{thm}
\Proof
To handle all these cases simultaneously, let us assume that $(G,m)$ is a $8$-target, and $v_1\d v_2\d v_3\d v_4\d v_5\d v_1$ are the vertices in order of some cycle of $G$,
and this cycle bounds a disc which is the union of three triangles of $G$, namely $v_1v_2v_3$, $v_1v_3v_5$ and $v_3v_4v_5$. Moreover, there is a 
subset $Z\subseteq \{v_1v_3,v_3v_5\}$ such that $m(e) = 0$ for all $e\in Z$ and deleting the edges in $Z$ gives a minimum $8$-counterexample. Finally, we assume that
$$m(v_1v_2) + m(v_1v_3)+m(v_2v_3)+m(v_3v_4)+m(v_3v_5)\ge 8,$$
and 
$$m^+(v_1v_2)+ m(v_1v_3)+m(v_3v_5)+m^+(v_4v_5)\ge 7.$$
To obtain the subcases Conf(11), Conf(12) and  Conf(13), we set, respectively,
\begin{itemize}
\item $Z = \{v_1v_3\}$, $m(v_1v_2)\ge 3$, $m(v_3v_4)\ge 3$, $m(v_3v_5) = 1$, $m^+(v_4v_5)\ge 3$, and $m(v_1v_5)\le 3$
\item $Z = \{v_3v_5\}$, $m^+(v_1v_2)\ge 3$, $m(v_2v_3) = 2$, $m(v_3v_4)\ge 2$, $m(v_1v_3) = 2$, $m(v_1v_5)\le 3$ and $m^+(v_4v_5)\ge 2$
\item $Z = \{v_1v_3,v_3v_5\}$, $m(v_1v_2)\ge \max(m(v_2v_3),m(v_1v_5))$.
\end{itemize}
(Edges not mentioned are unrestricted.)
Let $(G,m')$ be obtained by switching on the sequence $v_2\d v_3\d v_5\d v_1\d v_2$. (We postpone for the moment the question of whether this sequence is switchable.)
Let us suppose (for a contradiction) that $(G,m')$ admits an $8$-edge-colouring $F_1\l F_8$. Let $k = m(v_1v_2)+m(v_1v_3)+m(v_3v_5)+2$; then
we may assume that $F_1\l F_{k}$ each contain exactly one of $v_1v_2,v_1v_3,v_3v_5$, and $v_3v_5\in F_k$. Hence $k\le 8$.
Let $I = \{1\l 8\}$ if $m(v_3v_5)\ge 1$, and $I = \{1\l 8\}\setminus \{k\}$
otherwise. Since $v_2v_3$ meets all the edges $v_1v_2, v_1v_3,v_3v_5$, it follows that none of $F_1\l F_k$
contain $v_2v_3$, and so $k+m(v_2v_3)-1\le 8$ and we may assume that $v_2v_3\in F_j$ for $k+1\le j\le k+m(v_2v_3)-1$.  Thus there are exactly
$9-k-m(v_2v_3)$ values of $j\in \{1\l 8\}$ such that $F_j$ contains none of $v_1v_2, v_1v_3,v_3v_5, v_2v_3$. Since by hypothesis 
$$m(v_1v_2) + m(v_1v_3)+m(v_2v_3)+m(v_3v_4)+m(v_3v_5)\ge 8,$$
and so 
$m(v_3v_4) > 9-k-m(v_2v_3)$, there exists $h\le k+m(v_2v_3)-1$ such that $v_3v_4\in F_h$; since $v_3v_4$ meets each of $v_1v_3, v_2v_3$ and $v_3v_5$, 
it follows that $v_1v_2\in F_h$, and so $h< k$; and from the symmetry we may assume that $h = 1$.

For each $i\in I$ let $Q_i$ as in \ref{cuts}. Now $|F_j\cap Q_i|=1$ for $1\le j\le 8$ with $j\ne i$; and since $F_1$ contains $v_1v_2,v_3v_4$
it follows that for $i\ne 1$ $v_3v_4\notin Q_i$. Consequently $v_4v_5\in Q_i$ for all $i\in I\setminus \{1\}$. Let $r_1,r_2$ be the second regions for
$v_1v_2, v_4v_5$ respectively, and let their sets of doors be $D_1,D_2$.  Hence for each $j\in\{1\l 8\}$, since there exists $i\in I\setminus\{1\}$
with $i\ne j$, it follows that $F_j$ contains at most one of $v_1v_2,v_1v_3,v_3v_5,v_4v_5$, and so we may assume that $v_4v_5\in F_j$ for
$k+1\le j\le k'$ where $k' = k+m(v_4v_5)$, and in particular $k'\le 8$. From the hypothesis, $k'\ge 7$.
\\
\\
(1) {\em For $i \in I\setminus \{1\}$, one of $F_i\cap D_1$, $F_i\cap D_2$ is non-empty, and both if $k' = 8$ or $i=8$.}
\\
\\
Let $e_1\l e_n,e_1$ be the edges of $Q_i$ in order, where $e_1 = v_1v_2$, $e_2 = v_1v_3$, $e_3 = v_3v_5$ and $e_4 = v_4v_5$.
Thus for $1\le j\le k'$, $F_j$ contains one of $e_1, e_2, e_3, e_4$, and hence contains none of $e_5\l e_n$ if $j\ne i$. Now
since $F_i$ contains at most one of $e_1,e_2,e_3,e_4$ and $|F_i\cap Q_i|\ge 5$, it follows that $n\ge 8$. Hence $e_5\l e_n$
belong only to $F_i$, except that one belongs to $F_8$ if $i,k<8$. This proves (1) as usual.

\bigskip
Since $k'\le 8$, one of $r_1,r_2$ is small since 
$m^+(v_1v_2)+ m(v_1v_3)+m(v_3v_5)+m^+(v_4v_5)\ge 7$.
Consequently, (1) implies that $k'=7$; and so $r_1,r_2$ are both small, again a contradiction to (1). 

This proves that $(G,m')$ is not $8$-edge-colourable, and in particular the sequence $v_2\d v_3\d v_5\d v_1\d v_2$ is not switchable.
Let us look at the subcases for Conf(11), Conf(12), Conf(13) listed above. In the Conf(11) subcase,  
$m(v_1v_2)\ge 3\ge m(v_1v_5)$, so we only need to check that $m(v_1v_2)\ge m(v_2v_3)$. If not, then $m(v_2v_3) = 4$, contrary to Conf(2).
In the Conf(13) subcase, the condition that $m(v_1v_2)\ge \max(m(v_2v_3),m(v_1v_5))$ is explicitly given.
In the Conf(12) subcase, $m(v_1v_2)\ge 2\ge m(v_2v_3)$, so we only need to check that $m(v_1v_2)\ge m(v_1v_5)$. Suppose not; then 
$m(v_1v_5) = 3$ and $m(v_1v_2) = 2$. In this case the sequence $v_2\d v_3\d v_5\d v_1\d v_2$ is not switchable, so we need a different approach.

Since $(G,m')$ given above is not $8$-colourable, it follows from \ref{counterex} that $m'(\delta(X))\ge 10$ for every subset $X\subseteq V(G)$ with $|X|$ odd
and $|X|, |V(G)\setminus X|\ge 3$. Let $(G,m'')$ be obtained from $(G,m')$ by switching again on the same sequence. Now $(G,m'')$ is a $8$-target,
since $m(v_2v_3),m(v_1v_5)\ge 2$; and it is smaller than $(G,m)$, and therefore admits an $8$-edge-colouring, say $F_1\l F_8$. Since 
$m''(v_1v_2)+m''(v_1v_3)+m''(v_3v_5)+m''(v_1v_5) >8$, some $F_i$ contains two of $v_1v_2,v_1v_3,v_3v_5,v_1v_5$, and therefore
contains $v_1v_2$ and $v_3v_5$. By replacing $F_i$ by $(F_\setminus \{v_1v_2,v_3v_5\})\cup \{v_2v_3,v_1v_5\}$ we therefore obtain an $8$-edge-colouring of
$(G,m')$, a contradiction. This proves \ref{5gon}.~\bbox

\begin{thm}\label{bigedge}
No minimum $8$-counterexample contains Conf(14).
\end{thm}
\Proof
Let $(G,m)$ be a minimum $8$-counterexample, and suppose that some edge $uv$
is incident with regions $r_1,r_2$ where $r_1$ has at most six doors disjoint from $uv$, and $m(uv)\ge 5$, and either $m(uv)\ge 6$ or $r_2$ is small.
By exchanging $r_1,r_2$ if necessary, we may assume that if $r_1,r_2$ are both small, then the 
length of $r_1$ is at least the length of $r_2$.
By \ref{dense}, we do not have Conf(3), so not both $r_1,r_2$ are triangles, and by \ref{touchtri}, 
if $m(uv)\ge 6$ then neither of $r_1,r_2$ is a triangle; so $r_1$ is not a triangle.
Let $x\d u\d v\d y$ be a path of $C_{r_1}$.
Since $m(e)\ge 5$, this path is
switchable; let $(G',m')$ be obtained from $(G,m)$ by switching on it, and let $F_1\l F_8$ be an $8$-edge-colouring of $(G',m')$. 
Let $k = m'(uv)+m'(xy)\ge 7$.
Let $I = \{1\l 8\}\setminus \{k\}$ if $x,y$ are non-adjacent in $G$, and $I = \{1\l 8\}$ if $xy\in E(G)$.
For $i\in I$, let $Q_i$ be as in \ref{cuts}. Since $Q_i$ contains both $uv,xy$ for each $i\in I$, it follows that for $1\le j\le 8$, $F_j$ contains at most one
of $uv,xy$. Thus
we may assume that $uv\in F_i$ for $1\le i\le m'(uv)$, and $xy\in F_i$ for $m'(uv)<i\le k$.
Thus $k\le 8$.
Let $D_1$ be the set of doors for $r_1$ that are disjoint from $e$, and let $D_2$ be the set of doors for $r_2$.
\\
\\
(1) {\em For each $i\in I$, one of $F_i\cap Q_i\cap D_1, F_i\cap Q_i\cap D_2$ is nonempty, and if $k=8$ or $i>k$ then both are nonempty.}
\\
\\
Let $i\in I$, and let the edges of $Q_i$ in order be $e_1\l e_n,e_1$, where $e_1 = uv$ and $e_2 = xy$. Since $|F_i\cap Q_i|\ge 5$ and $F_i$ contains at most one of
$e_1,e_2$, it follows that $n\ge 6$.
Suppose that $k = 8$. Then for $1\le j\le 8$, $F_j$ contains one of $e_1,e_2$; and hence for all $j\in \{1\l 8\}$ with $j\ne i$, $e_3\l e_n\notin F_j$. It follows
that $e_n,e_{n-1}$ belong only to $F_i$ and hence $e_n\in F_i\cap Q_i\cap D_2$. Since this holds for all $i\in I$, it follows that $|D_2|\ge |I|\ge 7$. Hence $r_2$
is big, and so by hypothesis, $m(uv)\ge 6$. Since $k=8$ it follows that $xy\notin E(G)$. Consequently $e_3$ is an edge of $C_{r_1}$, and since $e_3,e_4$ belong
only to $F_i$, it follows that $e_3$ is a door for $r_1$. But $e_3\ne ux,vy$ from the choice of the switchable path, and so $e_3\in F_i\cap Q_i\cap D_1$. 
Hence in this case (1) holds.

Thus we may assume that $k = 7$; and so $m(e) = 5$, and $r_2$ is small, and $xy\notin E(G)$, and 
$uv\in F_1\l F_6$, and $xy\in F_7$. Thus $I = \{1\l 6,8\}$.
If $i=8$, then since $uv,xy\in Q_i$ and $F_j$ contains one of $e_1,e_2$ for all $j\in \{1\l 7\}$, it follows as before that $e_3\in F_i\cap Q_i\cap D_1$ and 
$e_n\in F_i\cap Q_i\cap D_2$.
Thus we may assume that $i\le 6$. For $1\le j\le 8$
with $j\ne i$, $|F_j\cap Q_i| = 1$, and for $1\le j\le 7$, $F_j$ contains one of $e_1,e_2$. Hence $e_3\l e_n$ belong only to $F_i$ and to $F_8$, and only
one of them belongs to $F_8$. If neither of $e_n,e_{n-1}$ belong to $F_8$ then $e_n\in F_i\cap Q_i\cap D_2$ as required; so we assume that $F_8$ contains one
of $e_n,e_{n-1}$; and so $e_3\l e_{n-2}$ belong only to $F_i$. Since $n\ge 6$, it follows that $e_3\in F_i\cap Q_i\cap D_1$ as required. This proves (1).

\bigskip

If $k = 8$, then (1) implies that $|D_1|\ge 7$ as required. So we may assume that $k = 7$ and hence $m(e) = 5$ and $xy\notin E(G)$; and $r_2$ is small.
Suppose that there are three values of $i\in \{1\l 6\}$ such that $|F_i\cap D_1| = 1$ and $F_i\cap D_2 = \emptyset$,
say $i = 1,2,3$. Let $f_i\in F_i\cap D_1$ for $i = 1,2,3$, and we may assume that $f_3$ is between $f_1$ and $f_2$
in the path $C_{r_1}\setminus\{uv\}$. Choose $X\subseteq V(G')$ such that $\delta_{G'}(X) = Q_{3}$. Since only one edge of $C_{r_1}\setminus\{e\}$
belongs to $Q_3$, one of $f_1, f_2$ has both ends in $X$ and the other has both ends in $V(G')\setminus X$; say $f_1$ has both ends in $X$.
Let $Z$ be the set of edges of $G'$ with both ends in $X$. Thus $(F_1\cap Z)\cup (F_2\setminus Z)$ is a perfect matching, since  $e\in F_1\cap F_2$, and no other edge
of $\delta_{G'}(X)$ belongs to $F_1\cup F_2$; and similarly $(F_2\cap Z)\cup (F_1\setminus Z)$ is a perfect matching. Call them $F_1', F_2'$ respectively.
Then $F_1', F_2',F_3, F_4\l F_8$ form an $8$-edge-colouring of $(G', m')$, yet $f_1,f_2$ are the only edges of $D_1\cup D_2$ included in $F_1'\cup F_2'$,
and neither of them is in $F_2'$, contrary to (1).
Thus there are no three such values of $i$; and similarly  there are at most two such that $|F_i\cap D_2| = 1$ and $F_i\cap D_1 = \emptyset$.
Thus there are at least three values of $i\in I$ such that $|F_i\cap D_1|+|F_i\cap D_2|\ge 2$ (counting $i = 8$), and so $|D_1|+|D_2|\ge 10$.
But $|D_1|\le 6$ by hypothesis and $|D_2|\le 3$ since $r_2$ is small, a contradiction.
This proves \ref{bigedge}.~\bbox

\begin{thm}\label{3heavy}
No minimum $8$-counterexample contains Conf(15) or Conf(16).
\end{thm}
\Proof
To handle both at once, we assume that $(G,m)$ is an $8$-target with a region $r$, and $uv\in E(C_r)$, and $uvw$ is another region, satisfying:
\begin{itemize}
\item either $(G,m)$ is a minimum $8$-counterexample, or $m(uv) = 0$ and deleting $uv$ gives a minimum $8$-counterexample
\item $m(uv)+m^+(uw)\ge 4$
\item every edge of $C_r$ not incident with $u$ is $3$-heavy
\item $m(vw)\le m(uw)$, and the second edge of $C_r$ incident with $u$ has multiplicity at most $m(uw)$.
\end{itemize}
Note that while Conf(16) fits these conditions, some instances of Conf(15) may not, and we will handle them later.
Let the second neighbour of $u$ in $C$ be $t$.

By hypothesis, the path $t\d u\d w\d v$ is switchable; let $(G',m')$ be obtained from it by switching, and let $F_1\l F_8$ be an $8$-edge-colouring of it.
Let $k = m(uw)+m(uv)+2\ge 5$; then we may assume that $F_1\l F_{k-1}$ contain one of $uw,uv$, and $tv\in F_k$. Let $I = \{1\l 8\}$ if $tv\in E(G)$, and
$I = \{1\l 8\}\setminus \{k\}$ otherwise. For each $i\in I$ let $Q_i$ be as in \ref{cuts}. Thus each $Q_i$ contains all of $uw,uv,tv$, and so
no edge of $Q_i\setminus\{uw,uv,tv\}$ belongs to $F_j$ for any $j\ne i$ with $j\le k$. 
\\
\\
(1) {\em $k = 5$.}
\\
\\
For suppose that $k\ge 6$. Choose $i\in I\cap \{7,8\}$. Since $Q_i$ contains $uv,uw,tv$, it follows that $F_1\l F_6$ all contain an edge in $\{uv,uw,tv\}\cap Q_i$; and hence
no edge of $Q_i\setminus\{uv,uw,tv\}$ belongs to any of $F_1\l F_6$. Choose an edge $f$ of $C_r\setminus \{u,v\}$ with $f\in Q_i$. Now $f\ne tu$ by the choice of
switchable path, and so $f$ is $3$-heavy (with respect to $(G,m)$), and if $f = tv$ then $m'(f)>m(f)$. 
Consequently there are three values of $j\in \{1\l 8\}\setminus \{k\}$ such that $F_j\cap Q_i$ contains an edge different from $uv,uw$, and hence some such $j$ belongs to 
$\{1\l 5\}$,
a contradiction. This proves (1).

\bigskip

Let $r_1$ be the second region for $uw$, and let $D_1$ be the set of doors for $r_1$.
From (1) it follows that $r_1$ is small, and so $|D_1|\le 3$.
\\
\\
(2) {\em For $i = 6,7,8$, $|Q_i\cap F_i\cap D_1| = 1$; and the edges of $F_6$ and $F_8$ in $Q_7$ have a common end (they may be the same).}
\\
\\
For let $i\in \{6,7,8\}$; then $i\in I$. Let the edges of $Q_i$ be $e_1\l e_n,e_1$ in order, where
$e_1 = uw$, $e_2 = uv$ and $e_3 = tv$. Then $n\ge 7$, since $|F_i\cap Q_i|\ge 5$. Let $h = 3$ if $tv\in E(G)$, and $h = 4$ otherwise.
Then $e_h$ is an edge of $C_r$ not incident with $u$, and so it
is $3$-heavy; and hence either $m(e_h) \ge 3$, or the second region for $e_h$ is a triangle and $e_{h+1}$ is an edge of it, and $m(e_h)+m(e_{h+1})\ge 3$.
Moreover, if $e_h = tv$ then $m'(e_h)>m(e_h)$. Thus in all cases it follows that
there are three values of $j\ne 5$
with $1\le j\le 8$ such that $F_j\cap Q_i$ contains one of $e_h,e_{h+1}$.
We deduce that these three values of $j$ are $6,7,8$, since $F_j\cap Q_i\subseteq \{uv,uw\}$ for $1\le j\le 4$.
Consequently for $1\le j\le 8$, $F_j\cap Q_i$ includes one of $e_1,e_2,e_3,e_4,e_5$. It follows that only $F_i$ contains $e_n,e_{n-1}$, and 
consequently $e_n\in Q_i\cap F_i\cap D_1$. Since $|D_1| = 3$, this proves the first assertion of (2). The second follows since, taking $i = 7$ and defining
$e_h$ as before, $F_6$ and $F_8$ each contain one of $e_h, e_{h+1}$, and these edges have a common end. This proves (2).

\bigskip

Let $Q_i\cap F_i\cap D_1 =\{f_i\}$ for $i = 6,7,8$. We may assume that $f_6,f_7,f_8$ are in order in the path $C_{r_1}\setminus \{uw\}$.
Choose $X\subseteq V(G)$ with $\delta_{G'}(X) = Q_7$.
Let $H$ be the subgraph of $G'$ with vertex set $V(G)$ and edge set $(F_6\setminus F_8)\cap (F_8\setminus F_6)$. Thus each component of $H$ is either a single vertex
or a cycle of even length. Now there are either no edges, or two edges, of $H$ that belong to $\delta_{G'}(X)$; and
if there are two then they have a common end by (2). It follows that the component of $H$, say $C$,
that contains $f_6$ does not contain $f_8$. Let $F_6' = (F_8\cap E(C))\cup (F_6\setminus E(C))$ and $F_8' = (F_6\cap E(C))\cup (F_8\setminus E(C))$; then
$F_6', F_8'$ are perfect matchings of $G'$, and $F_1\l F_5, F_6', F_7, F_8'$ is an $8$-edge-colouring of $(G,m')$. On the other hand both $f_6,f_8$ belong to
$F_8'$, so this $8$-edge-colouring does not satisfy (2), a contradiction. 

\bigskip

It remains to deal with the case of Conf(15) when the path $t\d u\d w\d v$ is not switchable. Thus, now we assume that 
\begin{itemize}
\item $(G,m)$ is a minimum $8$-counterexample
\item $r$ is a region of length at least four, and $e$ is an edge of $C_r$
\item $m^+(e)\ge 4$, 
and every edge of $C_r$ disjoint from $e$ is $3$-heavy
\item one of the edges of $C_r$ incident with $e$ has multiplicity more than $m(e)$.
\end{itemize}

Let $C_r$ have vertices $v_1\l v_p$ in order, where $p\ge 4$, $e = v_1v_2$, and $m(v_2v_3)>m(e)$.
It follows that $m(v_1v_2) = 3$ and $m(v_2v_3) = 4$. By \ref{dense},  we do not have Conf(4) so $p\ge 5$. The path $v_1\d v_2\d v_3\d v_4$
is switchable; let $(G,m')$ be obtained by switching on it. We may assume that $v_2v_3\in F_i$ for $1\le i\le 5$ and $v_1v_4\in F_6$. Since
$m'(v_1v_2) = 2$ and $v_1v_2$ meets both $v_2v_3$ and $v_1v_4$, it follows that $v_1v_2\in F_7,F_8$. Consequently $v_pv_1\in F_h$ for some $h$ with $1\le h\le 5$.
Let $I = \{1\l 8\}\setminus \{6\}$. For each $i\in I$ let $Q_i$ be as in \ref{cuts}.
Now $Q_7$ contains $v_2v_3,v_1v_4$, and so for $1\le j\le 6$, $F_j\cap Q_7\subseteq \{v_2v_3,v_1v_4\}$. In particular $v_pv_1\notin Q_7$. But $Q_7$ contains an edge $f$ of $C_r$, different from
$v_1v_2$, and this edge is $3$-heavy, since it is different from $v_pv_1$ and hence disjoint from $e$; 
and so $F_j\cap Q_i\setminus \{v_2v_3,v_1v_4\}\ne \emptyset$ for three values of $j\in \{1\l 8\}$, a contradiction.
This proves \ref{3heavy}.

\begin{thm}\label{not3heavy}
No minimum $8$-counterexample contains Conf(17) or Conf(18).
\end{thm}
\Proof
To handle both at once, we assume that $(G,m)$ is an $8$-target with a region $r$ with length at least four, and $uv\in E(C_r)$, and $uvw$ is another region, satisfying:
\begin{itemize}
\item either $(G,m)$ is a minimum $8$-counterexample, or $m(uv) = 0$ and deleting $uv$ gives a minimum $8$-counterexample
\item $m(uv)+m^+(uw)\ge 5$
\item let $t,x$ be the second neighbours of $u,v$ in $C_r$ respectively; if $m(uv) = 3$ and $e$ is $5$-heavy let $P = C_r\setminus \{u,v\}$, 
and otherwise let $P = C_r\setminus \{u\}$; then every edge $f$ of $P$ satisfies $m^+(f)\ge 2$, and at most one edge of $P$ is not $3$-heavy
\item $m(tu),m(vw)\le m(uw)$.
\end{itemize}
The path $t\d u\d w\d v$ is switchable; let $(G',m')$ be obtained by switching on it, and let
$F_1\l F_8$ be an $8$-edge-colouring of $(G', m')$. Since $r$ has length at least four, $tv\notin E(G)$. Let $k = m(uw)+(uv)+2\ge 6$; we may assume that
$F_i$ contains one of $uv,uw$ for $1\le i<k$, and $F_k$ contains $tv$. Let $I = \{1\l 8\}\setminus \{k\}$; and for each $i\in I$ let $Q_i$ be as in
\ref{cuts}. 
\\
\\
(1) {\em There is at most one value of $i\in I$ such that $Q_i\cap E(P)=\emptyset$, and if $i$ is such a value then $k = 7$ and $m(uv) = 3$ and $m(uw),m(vw) = 2$
and $uw\in F_i$.}
\\
\\
For suppose that $i\in I$ and $Q_i\cap E(P)=\emptyset$. It follows that $P = C_r\setminus \{u,v\}$, and so $m(uv) = 3$ and $m(uw),m(vw) = 2$, and $k = 7$. 
Now for $1\le i\le 7$, $F_i$ contains one of $uw,uv,tv$, and since $vw$ meets all of these edges it follows that $vw\in F_8$. 
But $vx$ belongs to some $F_j$ such that $F_j$ contains none of $tv,uv,vw$, and so $uw\in F_j$.
Then $|F_j\cap Q_i|\ge 2$, so $j = i$ and hence $uw\in F_i$. This proves (1).

\bigskip
Let $I'$ be the set of $i\in I$ such that $Q_i\cap E(P)\ne \emptyset$. By (1), $|I'|\ge 6$. Let $r_1$ be the second region for $uw$, and let its set of doors be $D_1$.
Thus $|D_1|\le 3$ if $k = 6$, since $m(uv)+m^+(uw)\ge 5$. Let $I''$ be the set of $i\in I'$ such that the edge in $Q_i\cap E(P)$ is not $3$-heavy.
\\
\\
(2){\em  There is a unique edge $f\in E(P)$ that is not $3$-heavy, and it belongs to none of $F_1\l F_k$.
Moreover, if $i\in I'\setminus I''$ then $k = 6$ and $i\le 5$ and $F_i\cap Q_i\cap D_1\ne \emptyset$.}
\\
\\
Suppose that $i\in I'\setminus I''$. There are therefore three values of $j\in \{1\l 8\}$ such that
$F_j\cap Q_i\not\subseteq \{uw,uv,tv\}$, and so at least two that are also different from $i$. Consequently, for those two
values of $j$, it follows that $uw,uv,tv\notin F_j$ and hence $k = 6$ and $j\in \{7,8\}$. Thus $i\le 5$. Let the edges of $Q_i$ in order be
$e_1\l e_n, e_1$, where $e_1 = uw$, $e_2 =uv$ and $e_3 = tv$; then $n\ge 7$, since $|F_i\cap Q_i|\ge 5$. But $F_1\l F_8$ each contain one of $e_1\l e_5$, so
$e_n\in F_i\cap Q_i\cap D_1$. This proves the second assertion of (2). For the first assertion, since $|D_1|\le 3$, 
it follows that $|I'\setminus I''|\le 3$. Since $|I'|\ge 6$, it follows
that $|I''|\ge 3$. But by hypothesis, there is at most one edge 
in $P$ that is not $3$-heavy, and so this edge exists, say $f$. It follows that $f\in Q_i$, for all $i\in I''$.
Now let $j\in \{1\l k\}$. Choose $i\in I''$
with $i\ne j$; then $F_j\cap Q_i\subseteq \{uw,uv,tv\}$, and so $F_j$ does not contain $f$. This proves (2).

\bigskip
By (2) we may assume that $f\in F_{k+1}$. Let $r_2$ be the second region at $f$, and let $D_2$ be its set of doors. By hypothesis, if $m(f) = 1$ then $|D_2|\le 3$.

Suppose that $k\ge 7$. By (2), $I'' = I'$ and $m(f) = 1$.
Let $i\in I'$, and let the edges of $Q_i$ in order be
$e_1\l e_n$, where $e_1 = uw$, $e_2 =uv$, $e_3 = tv$, and $e_4=f$. Since only one of $e_1\l e_4$ belongs to $F_i$, and $|F_i\cap Q_i|\ge 5$,
it follows that $n\ge 8$. But $F_1\l F_{8}$
each contain one of $e_1\l e_4$, and so $e_5\l e_n$ only belong to $F_i$; and hence $e_5\in F_i\cap Q_i\cap D_2$. Consequently $|D_2|\ge |I'|\ge 6$,
a contradiction. 

This proves that $k =6$, and hence $|D_1|\le 3$, and $I' = I$ by (1), and $7,8\in I''$ by (2). Now let $i\in I''$. Let the edges of $Q_i$ in order be
$e_1\l e_n, e_1$, where $e_1 = uw$, $e_2 =uv$, $e_3 = tv$, and $e_4=f$. Again $n\ge 8$.

Suppose that $m(f)\ge 2$; then $m(f) = 2$ by (2), and $f\in F_7,F_8$, and so $F_1\l F_8$ 
each contain one of $e_1\l e_4$, and therefore $e_5\l e_n$ belong to no $F_j$ with $j\ne i$. Since $n\ge 8$, it follows that $e_n\in D_1$, and so
$F_i\cap Q_i\cap D_1\ne \emptyset$. By (2), it follows that $F_i\cap Q_i\cap D_1\ne \emptyset$ for all $i\in I'$, and so $|D_1|\ge |I'| = 7$, a contradiction.
Thus $m(f) = 1$, and so $|D_2|\le 3$.

Again, let $i\in I''$, and let $e_1\l e_n, e_1$ be as before. Now $F_1\l F_{7}$
each contain one of $e_1\l e_4$, and so $e_5\l e_n$ belong to no $F_j$ with $1\le j\le 7$ and $j\ne i$, and only one of them belongs to $F_8$ if $i\ne 8$.
We assume first that $i\ne 8$. Since $n\ge 8$, either $e_5,e_6\notin F_8$, or $e_n,e_{n-1}\notin F_8$, and so either $e_5\in D_2$ or $e_n\in D_1$.
Now we assume $i = 8$. Then $e_5\l e_n$ belong to no $F_j$ with $1\le j\le 7$, and so $e_5\in D_2$ and $e_n\in D_1$.

In summary, we have shown that for each $i\in I''$, either $F_i\cap D_1\ne \emptyset$, or $F_i\cap D_2\ne \emptyset$ (both if $i = 8$); and $8\in I''$.
By (2), if $i\in I'\setminus I''$ then either  $F_i\cap D_1\ne \emptyset$, or $F_i\cap D_2\ne \emptyset$; and so
$|D_1|+|D_2|\ge |I'|+1\ge 7$, a contradiction. This proves \ref{not3heavy}.~\bbox

\begin{thm}\label{lastone}
No minimum $8$-counterexample contains Conf(19).
\end{thm}
\Proof
Let $(G,m)$ be a minimum $8$-counterexample, and suppose that $r$ is a region with length at least five, and $e$ is an edge of $C_r$, such that $m^+(e)\ge 5$, and
every edge of $C_r$ disjoint from $e$ is $2$-heavy, and at most two of them are not $3$-heavy. 
By \ref{not3heavy}, we do not have Conf(17), so
there are at least two edges in $C_r$ disjoint from $e$ that are not $3$-heavy, and so by hypothesis, there are exactly two, say
$g_1,g_2$. Thus $m(g_1),m(g_2)\le 2$. By hypothesis, $g_1,g_2$ are $2$-heavy.

Let $e=uv$, and let the second neighbours of $u,v$ in $C_r$ be $t,w$ respectively. Since $m(e) \ge 4$, it follows
that $m(tu),m(vw)\le m(uv)$ and so the path $t\d u\d v\d w$ is switchable. Let $(G',m')$ be obtained by switching on this path, and let $F_1\l F_8$
be an $8$-edge-colouring of it. Let $k = m(e)+2$. We may assume that $tw\in F_k$.
Let $I = \{1\l 8\}\setminus \{k\}$, and for each $i\in I$ let $Q_i$ be as in \ref{cuts}. Let $I_1,I_2,I_3$ be the sets of $i\in I$ such that
$g_1\in Q_i$, $g_2\in Q_i$, and $g_1,g_2\notin Q_i$ respectively.
\\
\\
(1) {\em $k = 6$.}
\\
\\
For suppose that $k>6$. Let $i\in I$, and 
let the edges of $Q_i$ in order be $e_1\l e_n, e_1$, where
$e_1 = uv$ and $e_2 = tw$. Thus $e_3$ is an edge of $C_r$ disjoint from $e$. 
Since $|F_i\cap Q_i|\ge 5$ and $|F_i\cap \{e_1,e_2\}|\le 1$, it follows that $n\ge 6$. 
Now there are $k\ge 7$ values of $j\in \{1\l 8\}$ such that $F_j$ contains one of $e_1,e_2$; and so there is at most value of $j\ne i$ such that $F_j$ contains one of $e_3,e_4$.
It follows that $e_3$ is not $3$-heavy and so $i\in I_1\cup I_2$. Since this holds for all $i\in I$, we may assume that $|I_1|\ge 4$. Let $i\in I_1$; as before,
there is at most one value of $j\ne i$ such that $F_j$ contains one of $e_3,e_4$. Now $m(g_1)\le 2$. If $m(g_1) = 2$, then
$g_1\in F_i$, and since this holds for all $i\in I_1$ it follows that $g_1$ is contained in $F_i$ for four different values of $i$, a contradiction. Thus
$m(g_1) = 1$. Since $g_1$ is $2$-heavy, the second region for $g_1$ is a triangle with edge set $\{g_1,p,q\}$ say, where $e_4 = p$. Hence one of
$f_1,p,q$ has multiplicity one and is contained in $F_i$. Since this holds for all $i\in I_1$ and $|I_1|\ge 4$, this is impossible. This proves (1).

\bigskip

We may therefore assume that $uv\in F_i$ for $1\le i\le 5$
and $tw\in F_6$. Since $k = 6$, it follows that $m(e) = 4$ and since $m^+(e)\ge 5$, the second region $r_1$ for $uv$ is small. Let $D_1$
be its set of doors.
\\
\\
(2) {\em If $i\in I_3$ then $i\le 5$ and $F_i\cap Q_i\cap D_1\ne \emptyset$.}
\\
\\
For let the edges of $Q_i$ in order be $e_1\l e_n, e_1$, where
$e_1 = uv$ and $e_2 = tw$. Then $F_1\l F_6$ each contain an edge in $\{e_1,e_2\}$, and so for $1\le j\le 6$ with $j\ne i$, none of $e_3\l e_n$
belongs to $F_j$. Now $e_3$ is $3$-heavy, and so there are three values of $j$ such that $F_j$ contains one of $e_3,e_4$; and so these three values are
$i,7,8$, and $i\ne 7,8$. (Thus $i\le 5$ since $6\notin I$.)
Hence for $1\le j\le 8$, $F_j$ contains one of $e_1\l e_4$; and so $e_n,e_{n-1}$ belong only to $F_i$. Hence $e_n \in D_1$. This proves (2).

\bigskip
For $j = 1,2$, let $I_j'$ be the set of all $i\in I_j$ such that $F_i\cap Q_i\cap D_1= \emptyset$.
\\
\\
(3) {\em For $j = 1,2$, $|I_j'|\le 2$, and $7,8\notin I_j'$, and if $|I_j'|=2$ then  $7,8\notin I_j$.}
\\
\\
For let $j = 1$ say. Suppose first that $m(g_1) = 2$, and let $g_1\in F_a,F_b$ where $1\le a<b\le 8$. 
Let $i\in I_1'$, and let $e_1\l e_n$ be as before; then $e_3=g_1$. 
Again, for $1\le j\le 6$ with $j\ne i$, none of $e_3\l e_n$
belong to $F_j$, and consequently $a,b \in \{i,7,8\}$. In particular, $b\ge 7$, and $a\in \{i,7\}$. Thus if $a\le 6$ then $i=a$ and so $|I_1'| = 1$ and the claim holds.
We assume then that $(a,b) = (7,8)$. But then $F_1\l F_8$ each contain one of $e_1,e_2,e_3$, and so $e_n\in D_1$, contradicting that $i\in I_1'$.
So the claim holds if $m(g_1) = 2$.

Next we assume that $m(g_1) = 1$. Since $g_1$ is $2$-heavy, the second region at $g_1$ is a triangle with edge set $\{g_1,p,q\}$ say. Let $g_1\in F_a$.
Let $i\in I_1'$, and let $e_1\l e_n$ be as before; then $e_3=g_1$. 
Again, for $1\le j\le 6$ with $j\ne i$, none of $e_3\l e_n$
belongs to $F_j$, and consequently $a \in \{i,7,8\}$. Thus if $a\ne 7,8$ then $i = a$ and $|I_1'| = 1$ and the claim holds. We assume then that $a=7$.
Thus each of $F_1\l F_7$ contains one of $e_1,e_2,e_3$, and for $1\le j\le 7$ with $j\ne i$, $F_j$ contains none of $e_4\l e_n$. 
Since $F_i\cap Q_i\cap D_1= \emptyset$, there exists $j\in \{1\l 8\}$ with $j\ne i$ such that $F_j$ contains one of $e_n,e_{n-1}$; and hence $j = 8$,
and so $i\ne 8$. (Also, $i\ne 7$ since $g_1\in F_7$ and $g_1$ meets $e_4$. Consequently, $7,8\notin I_j'$.)
Thus $F_1\l F_8$ each contain one of $e_1,e_2,e_3,e_{n-1},e_n$, and so $e_4$ is only contained in $F_i$. Consequently, $i$ has the property
that one of $p,q$ has multiplicity one, and $F_i$ contains it. Thus there are at most two such values of $i$, and so $|I_j'|\le 2$. Moreover,
if there are two such values, say $c,d$, then $c,d\le 5$ and $F_c$ contains one of $p,q$ and $F_d$ contains the other. Consequently if $7\in I_1$,
then one of $F_c,F_d$ contains two edges of $Q_7$, a contradiction. So if $|I_j'|=2$ then  $7,8\notin I_j$. This proves (3).

\bigskip
From (2), we may assume that $7\in I_1$, and so $|I_1'|+|I_2'|\le 3$ by (3). Consequently there are at least four values of $i\in I$ such that
 $F_i\cap Q_i\cap D_1\ne \emptyset$, and so $|D_1|\ge 4$, a contradiction. This proves \ref{lastone}.~\bbox

\bigskip

This completes the proof of \ref{reduc} and hence of \ref{mainthm}.
Perhaps despite appearances, there was some system to our choice of the $\beta$- and $\gamma$-rules. We started with the idea that we would normally pass a charge of one
from each small region to each big region sharing an edge with it, and made the minimum modifications we could to the $\beta$-rules so that
the proof of \ref{bigovercharge} worked. Then we experimented with the $\gamma$-rules to make \ref{triovercharge1}, \ref{triovercharge2} and \ref{smallovercharge}
work out.

It is to be hoped that solving these special cases of the main conjecture \ref{mainconj} will lead us to a proof of the general case, but that seems far away at the moment.
The same approach does indeed work (more simply) for seven-regular planar graphs, and this gives an alternative proof of the result of \cite{katie}, to appear in \cite{kawa}. We tried
the same again for nine-regular graphs, but there appeared to be some serious difficulties. Maybe more perseverance will bring it through, but it seems much harder than
the eight-regular case.


\begin{thebibliography}{99}
\bibitem{appelhaken1} K.Appel and A.Haken, ``Every planar map is four colorable. Part I. Discharging'', {\em Illinois J. Math.} 21 (1977), 429--490.
\bibitem{appelhaken2} K.Appel, A.Haken and J.Koch, ``Every planar map is four colorable. Part II. Reducibility'', {\em Illinois J. Math.} 21 (1977), 491--567.
\bibitem{dvorak} Z.Dvorak, K.Kawarabayashi and D.Kral, ``Packing six $T$-joins in plane graphs'', manuscript (2010arXiv1009.5912D)
\bibitem{katie} K.Edwards, {\em Optimization and Packings of $T$-joins and $T$-cuts}, M.Sc. Thesis, McGill University, 2011.
\bibitem{guenin} B.Guenin, ``Packing $T$-joins and edge-colouring in planar graphs'', {\em Mathematics of Operations Res.}, to appear.
\bibitem{kawa} M.Chudnovsky, K.Edwards, K.Kawarabayashi and P.Seymour, ``Edge-colouring seven-regular planar graphs'', in preparation.
\bibitem{rsst} N.Robertson, D.Sanders, P.Seymour and R.Thomas, 
``The four colour theorem'', {\em J. Combinatorial Theory, Ser. B}, 70 (1997), 2--44.
\bibitem{seymour} P. Seymour, {\em Matroids, Hypergraphs and the Max.-Flow 
Min.-Cut Theorem}, D.Phil. thesis, Oxford, 1975, page 34.
\bibitem{tait} P.G.Tait,  ``Remarks on the colourings of maps'', {\em Proc. R. Soc. Edinburgh} 10 (1880), 729.
\end{thebibliography}
\end{document}